\documentclass[twocolumn,tighten]{aastex63}

\usepackage{amsmath}
\usepackage{multirow}
\usepackage{ulem}
\usepackage{booktabs}

\begin{document}

\title{The Substructures in Disks undergoing Vertical Shear Instability:     \\ 
II. Observational Predictions for the Dust Continuum}

\author{Diana Blanco}
\affiliation{Department of Physics and Astronomy, California State University Northridge, 18111 Nordhoff Street, Northridge, CA 91330, USA}
\affiliation{Department of Astronomy and Astrophysics, University of California, Santa Cruz, 1156 High Street, Santa Cruz, CA 95064, USA}

\author{Luca Ricci}
\affiliation{Department of Physics and Astronomy, California State University Northridge, 18111 Nordhoff Street, Northridge, CA 91330, USA}

\author{Mario Flock}
\affiliation{Max-Planck-Institut f{\"u}r Astronomie, K{\"o}nigstuhl 17, D-69117 Heidelberg, Germany}

\author{Neal Turner}
\affiliation{Jet Propulsion Laboratory, California Institute of Technology, Pasadena, CA 91109, USA}

\correspondingauthor{Luca Ricci}
\email{luca.ricci@csun.edu}

\begin{abstract}
High-angular resolution observations at sub-millimeter/millimeter wavelengths of disks surrounding young stars have shown that their morphology is made of azimuthally-symmetric or point-symmetric substructures, in some cases with spiral arms, localized spur- or crescent-shaped features. The majority of theoretical studies with the aim of interpreting the observational results have focused on disk models with planets, under the assumption that the disk substructures are due to the disk-planet interaction. However, so far only in very few cases exoplanets have been detected in these systems. Furthermore, some substructures are expected to appear \textit{before} planets form, as they are necessary to drive the concentration of small solids which can lead to the formation of planetesimals.
In this work we present observational predictions from high-resolution 3D radiative hydrodynamical models which follow the evolution of gas and solids in a protoplanetary disk. 
We focus on substructures in the distribution of millimeter-sized and smaller solid particles produced by the vertical shear instability. We show that their characteristics are compatible with some of the shallow gaps detected in recent observations at sub-mm/mm wavelengths, and present predictions for future observations with better sensitivity and angular resolution with ALMA and a Next Generation Very Large Array. 
\end{abstract}

\section{Introduction \label{sec:intro}}

Planets form from the solid and gas material contained in disks orbiting young stars. The currently accepted theory for forming planets, the core accretion scenario, invokes the agglomeration of small dust particles into km-sized ``planetesimals'', which are massive enough to attract other solids in the disk \citep[for a review, see][]{Johansen:2014}. 
The distribution of solids with sizes comparable or smaller than $\sim$ 1-10 cm in the dense regions of the disk midplane can be investigated at (sub-)millimeter and radio wavelengths, where the optical depth of the dust emission is significantly lower than in the infrared. 
High angular resolution observations using the interferometric technique in this region of the electromagnetic spectrum resulted in the discovery of several morphological features, including rings and gaps, inner cavities, spirals arms, arcs, and crescents in the spatial distribution of the circumstellar material \citep[e.g.,][]{Casassus:2013,ALMAPartnership:2015, Dipierro:2018, Andrews:2018}.

Several of these substructures, especially the prominent rings, gaps and inner cavities, may originate from gravitational perturbations induced by yet unseen planets in the act of forming. Following this hypothesis, several authors have modeled the putative planet-disk interaction to match the observed disk substructures and produced estimates for the mass and orbital radius for
several supposed young planets \citep{Jin:2016,Liu:2018,Dipierro:2018}.

However, the planet hypothesis is not necessarily unique as other mechanisms may potentially produce similar substructures to those recently observed. For example, depending on the physical conditions, young disks are expected to be prone to a variety of hydrodynamical instabilities, with or without the interplay of a magnetic field, which may be effective at producing significant substructures in the distribution of gas and dust in disks \citep[see][for a recent review]{Lyra:2019}. Furthermore, spatial inhomogeneities in the distribution of solids are thought to be necessary to overcome the cm-to-m size barrier in models of planet formation \citep{Weidenschilling:1977,Brauer:2007,Brauer:2008}. As a consequence, disks with substructures which are \textit{not} due to planets are expected from our current understanding of planet formation.
Investigating the development, evolution and possible appearance of these structures via numerical simulations for the gas and dust in disks undergoing hydrodynamical instabilities is therefore very important to provide possible alternative ways to explain the observational results, as well as to produce predictions for future observations that may help disentangle the physical processes which are responsible for those structures. 

Deriving observational predictions from models with hydrodynamical instabilities is key also to tackle the long standing question of what is the physical mechanism that is responsible for the transport of angular momentum in the disk. 
In protoplanetary disks, the transport of angular momentum has been conventionally attributed to turbulence driven by the magneto-rotational instability \citep[MRI,][]{Balbus:1991}.
However, more recent theoretical studies have shown that, under the low ionization and consequent weak coupling between the gas and magnetic field as expected in young disks, magnetically driven instabilities can be strongly suppressed in large regions of the disk \citep{Turner:2014,Armitage:2015} due to the non-ideal MHD terms of Ohmic resistivity,
ambipolar diffusion, and Hall effect \citep{Wardle:2007,Salmeron:2008,Bai:2011, Bai:2014,Dzyurkevich:2013}. In such regions, hydrodynamical instabilities, which do not involve coupling with the magnetic field, may play an important role in driving the transport of angular momentum, and in general influence the dynamics of the gas, including turbulence.

\begin{figure*}[thb!]
\centering
\includegraphics[width=\textwidth]{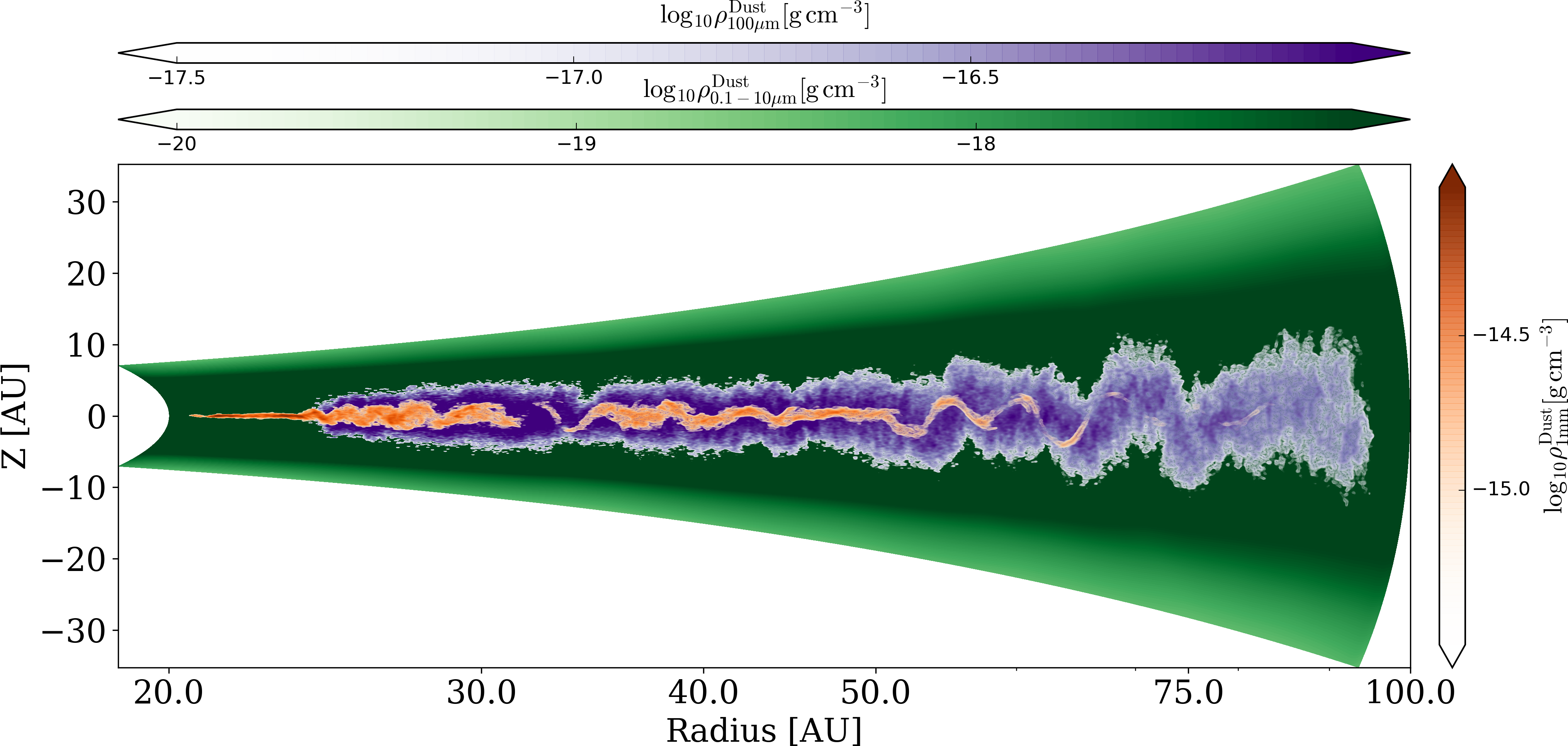}
\caption{Density of grains with sizes of $0.1 - 10~\micron$ (green color scale), $100~\micron$ (blue) and 1~mm (orange) on the radial-vertical plane after averaging over the azimuthal coordinate.}
\label{fig:dust_r_z}
\end{figure*}

In this work we investigate the observational predictions of 3D global radiative hydrodynamical models of disks which follow the dynamics of gas and solids. In these models, which follow the prescriptions described in \citet{Flock:2017, Flock:2020} to solve the radiative hydrodynamic equations using the
hybrid stellar irradiation and flux-limited diffusion method, a large fraction of the disk is subject to the vertical shear instability \citep[VSI,][]{Nelson:2013,Pfeil:2019}, which is one of the most promising hydrodynamical mechanisms proposed to drive turbulence in real disks. The VSI is the application of the Goldreich-Schubert-Fricke instability \citep{Goldreich:1967,Fricke:1968} to the case of accretion disks in which the rotational velocity of gas varies with vertical height from the disk midplane. The VSI acts in regions in which the cooling timescale is shorter than the orbital timescale to avoid the stabilizing effect of vertical buoyancy. Under these conditions, the VSI has been shown to develop into efficient turbulence with Shakura-Sunyaev $\alpha$ values up to $\sim 10^{-3}$ \citep{Shakura:1973,Manger:2020}, accompanied by strong vertical oscillations of the gas. Moreover, recent 3D simulations have shown that the VSI can also trigger the development of vortices, i.e. anticyclonic structures that
rotate in the plane of the disk, and have radial and azimuthal extents similar to, or a little bigger than the disk vertical thickness \citep{Richard:2016,Manger:2018,Latter:2018}. 

Besides affecting the dynamics of the gas, the VSI may have a strong impact also on the early stages of planet formation. The vertical oscillations of the gas can efficiently stir up the small dust particles \citep{Stoll:2016,Flock:2017}, whereas relatively large grains can get trapped in the vortices, which may facilitate planetesimal formation \citep[e.g.,][]{Manger:2018}.  
As a consequence, these models may predict signatures in the dust continuum emission of young disks which may be detectable through high-angular resolution interferometric observations at sub-millimeter/millimeter wavelengths. 
The characterization of these predictions for the dust continuum emission is the main goal of this work.  Given the condition of efficient cooling, relative to the local orbital timescale, the VSI is expected to play an important role at orbital radii outside $> 5$ au, corresponding to the disk regions in which ALMA observations have detected substructures in an increasing sample of nearby disks. We therefore discuss the results of our calculations in light of these recent observational findings. We then investigate the capabilities of a future Next Generation Very Large Array \citep[ngVLA,][]{Murphy:2018} to detect and spatially resolve these substructures down to unprecedented resolutions at millimeter and centimeter wavelengths. 

We also note that although the models discussed in this work are purely hydrodynamical, i.e. they neglect the effects of the possible coupling with a magnetic field, recent 2D global non-ideal MHD simulations with ambipolar diffusion and Ohmic resistivity suggest that the properties of the VSI remain similar to the unmagnetized case \citep{Cui:2019}.

Section~\ref{sec:sims} provides an introduction to the numerical simulations and the synthetic images extracted from the disk model, Section~\ref{sec:obs} describes the methods to simulate ALMA and ngVLA observations for the dust continuum emission of the model, Section~\ref{sec:results} illustrates the results of these simulated observations, Section~\ref{sec:discussion} presents a discussion of the main results of this work and Section~\ref{sec:conclusion} summarizes the conclusions.

\section{Hydrodynamical and Radiative Transfer Simulations} \label{sec:sims}

The disk simulations are computed using high-resolution 3D radiation hydrodynamical calculations which include irradiation by the star and embedded grains. In particular, the model described here is the same as presented in \citet{Flock:2020}.
The radiation hydrodynamic equations are solved using the same method as described in \citet{Flock:2017}. In that work, the numerical simulations were run with a reduced azimuthal domain for 400 orbits (at the inner radius of the radial domain of 20 au). The new simulations presented in this work are a continuation of the \citet{Flock:2017} simulations for 400 more orbits, but also extend the azimuthal domain to 2$\pi$. This is key to obtain reliable predictions for disk substructures which can be compared with observations.  

\begin{figure*}[thb!]
\centering
\includegraphics[width=0.85\textwidth]{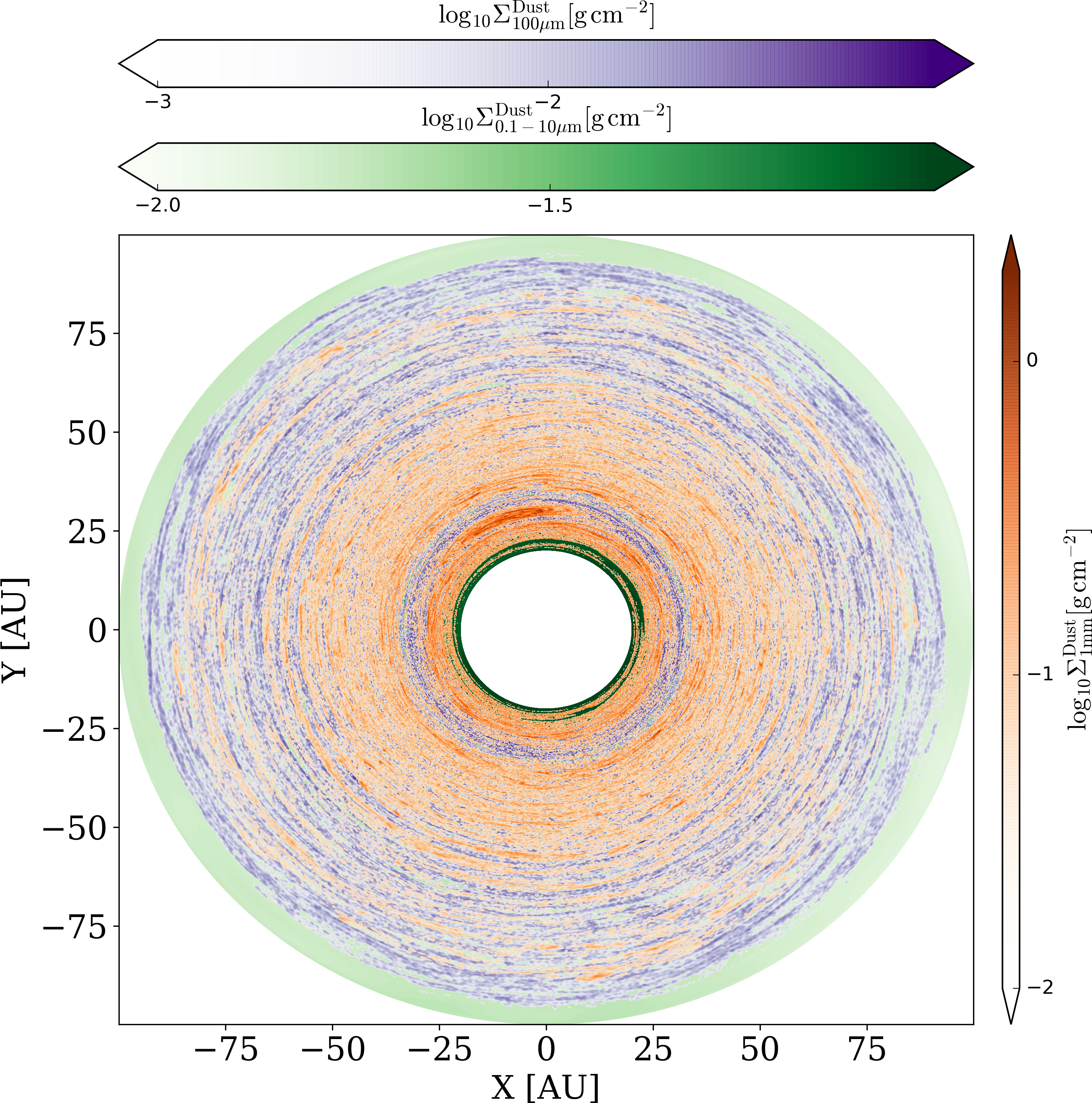}
\caption{2D surface density map of grains with sizes of $0.1 - 10~\micron$ (green color scale), $100~\micron$ (blue) and 1~mm (orange).}
\label{fig:dust_r_p}
\end{figure*}

The setup for the disk model discussed in this work was chosen based on the properties observed for disks around T Tauri stars. A summary of some of the main parameters for the model is shown in Table~\ref{tab:model}. The radial domain of our simulations is taken to be $20 - 100$ au, which covers the spatial region of the disk in which previous 3D global hydrodynamical simulations have seen the development of VSI instabilities. It is also in these disk regions that recent high-angular resolution imaging with ALMA has identified substructures in nearly all the observed disks \citep[e.g.,][]{Andrews:2018}. As a result, all the images presented in this work show a central cavity with stellocentric radius of 20 au. Since these regions are outside of the domain of our hydrodynamic simulations, we decided to mask them using a gray circle. We note here that the addition of bright emission from these inner disk regions would not imply any significant issue in the imaging related to dynamic range. In fact, by extrapolating the brightness of our models from a stellocentric radius of 20 au down to 1 au under the assumption of optically thick emission, we obtain dynamic range values in the simulated observed maps of about 120 at 0.87 mm and lower values at longer wavelengths.

We also note that the Rosseland mean opacity $\kappa_d$ and the Planck opacity for the irradiation $\kappa_*$ are used in the radiation hydrodynamical simulations to capture the irradiation heating and to solve for the heating and cooling in the disk. Even if these values are not fully self-consistent with the adopted opacities for the radiative transfer calculations (see Section~\ref{sec:synthetic}), they are good approximations to determine the main heating and cooling of the disk, and the discrepancy would not have any significant effect on the results presented in this work.
For more details on the other parameters of the model, main assumptions, and procedure to solve the hydrodynamic equations we refer to \citet{Flock:2017,Flock:2020}.

In this simulation, the dust component is included in the simulation after 250 inner orbits and is divided into 3 bins of grain sizes. The first bin represents the small grains with sizes of $0.1 - 10~\mu$m distributed with a power-law grain size distribution with exponent $-3.5$ (Eq.~\ref{eq:grainsizedist}), which are small enough to be treated as fully coupled to the dynamics of the gas in our simulations. The other two bins represent larger grains with single dust grain sizes of $\rm{100~\mu}$m and $\rm{1~mm}$, respectively, which are less coupled to the gas motion. These grains are individually evolved in the simulation with a size-dependent gas drag. We instead do not include in this work the effects of the dust back-reaction on the gas. We include $5 \times 10^5$ particles for each of the two bins of larger grains. More information on the procedure adopted to divide the surface density of solid particles among the 3 size bins is provided in Appendix~\ref{sec:appendix_dust_bins}.

In the following we present the results of our disk model and simulations of observations calculated at 700 inner orbits after the beginning of the numerical simulation. Although the density distribution of gas and dust do vary somewhat between the 400 and 800 total inner orbits spanned by our simulations, we do not investigate here the time evolution of these structures. A detailed discussion on this evolution is presented in \citet{Flock:2020}. 

\begin{table}
\begin{tabular}{p{3cm}p{4.7cm}}
  \hline
  \hline
Surface density & $\rm \Sigma=6.0 \left ( \frac{r}{100 AU} \right )^{-1} g/cm^2$ \\
Stellar parameters & $\rm T_{\rm{eff},\star}=4000\, K$,\\ & $\rm R_{\star}=2.0\, R_\sun,\, M_{\star}=0.5\, M_\sun$\\
Opacities & $\rm \kappa_*=1300\, cm^2/g$\\
        & $\rm \kappa_d = 400\, cm^2/g$\\
  Dust-to-gas ratio  & $D2G=10^{-2}$ \\    
  
  Setup         & Grid:  $r \, : \, \theta \, : \, \phi$\\
                & Domain: $\rm 20-100~au\, : \, \pm 0.35~rad: \, 2\pi$ rad\\
		& Resolution: $1024 \times 512 \times 2044$\\
  \hline
  \hline
\end{tabular}
\caption{Setup parameters for the 3D radiation HD disk model, including the
  gas surface density, stellar parameters, Planck and Rosseland mean opacity values, the total dust to gas mass ratio, the domain and resolution of the numerical simulations. The domain for the polar angle $\theta$ is expressed in terms of the angles above and below the disk midplane and centered on the stellar position.}
\label{tab:model}
\end{table}

\subsection{Density structure of the disk model}

Figure~\ref{fig:dust_r_z} displays the density distribution of the three different dust size bins in the vertical and radial directions of the disk after averaging over the azimuthal coordinate. As expected from the size-dependent coupling between dust and gas, the smallest grains with sizes of $0.1 - 10~\micron$ share the same spatial distribution as the gas, whereas the larger grains with sizes of $100~\micron$ and $1$ mm are more settled towards the disk midplane. The effects of the characteristic vertical upward and downward motions triggered by the VSI are clearly visible in the figure \citep{Flock:2017,Flock:2020}, and are more pronounced for the mm-sized grains which have a vertically thinner distribution.

Regarding the radial distribution of dust, the mm-sized grains are confined to smaller radii than the 0.1-mm sized grains as a consequence of their faster inward radial drift. Moreover, the mm-sized grains show pronounced substructures also in the radial direction, which manifest themselves as a succession of overdense and underdense regions in Fig.~\ref{fig:dust_r_z}, with the highest contrast found at $30 - 35$ au from the star.

A face-on view on the dust surface density for each bin size is presented in Fig.~\ref{fig:dust_r_p}. Here, the substructure previously identified in the $30 - 35$ au region shows itself as a pronounced annular gap in the distribution of mm-sized grains. Towards the north side of the inner edge of the gap, a strong local concentration of mm-sized grains is visible. This is due to a large vortex triggered by the combination of VSI and Rossby Wave Instability (RWI), as described in Section~\ref{sec:discussion}. In the disk regions further from the star, both elongated arc-like structures and more compact clumps of mm grains are present, which are due to short lived vortices. The nature and evolution of these short lived vortices and emerging substructures are discussed in \citet{Flock:2020}.


\subsection{Disk synthetic images for the dust continuum emission at sub-mm and cm wavelengths \label{sec:synthetic}}

To post-process the simulation data we use the publicly available Monte-Carlo radiation transfer code RADMC-3D, which can treat dust having more than one size distribution \citep{Dullemond:2012}. The opacities for the three size bins in our simulations were calculated with Mie theory using the tool MieX \citep{Wolf:2004}, adopting the optical data from \citet{Weingartner:2001}. For the dust composition we assumed composite grains made of silicate, perpendicular and parallel graphite with volume fractions of 62.5, 25, 12.5$\%$, respectively \citep[see Appendix C in][]{Flock:2017}. The derived values for the absorption and scattering dust opacity coefficients $k_{\rm{abs}}$ and $k_{\rm{sca}}$, respectively, for the three grain size bins at the wavelengths of the model images are listed in Table~\ref{table:opacities}.

\begin{figure*}[t!]
\centering
\includegraphics[height=22cm, trim = {1.5cm 2cm 1.5cm 12cm}]{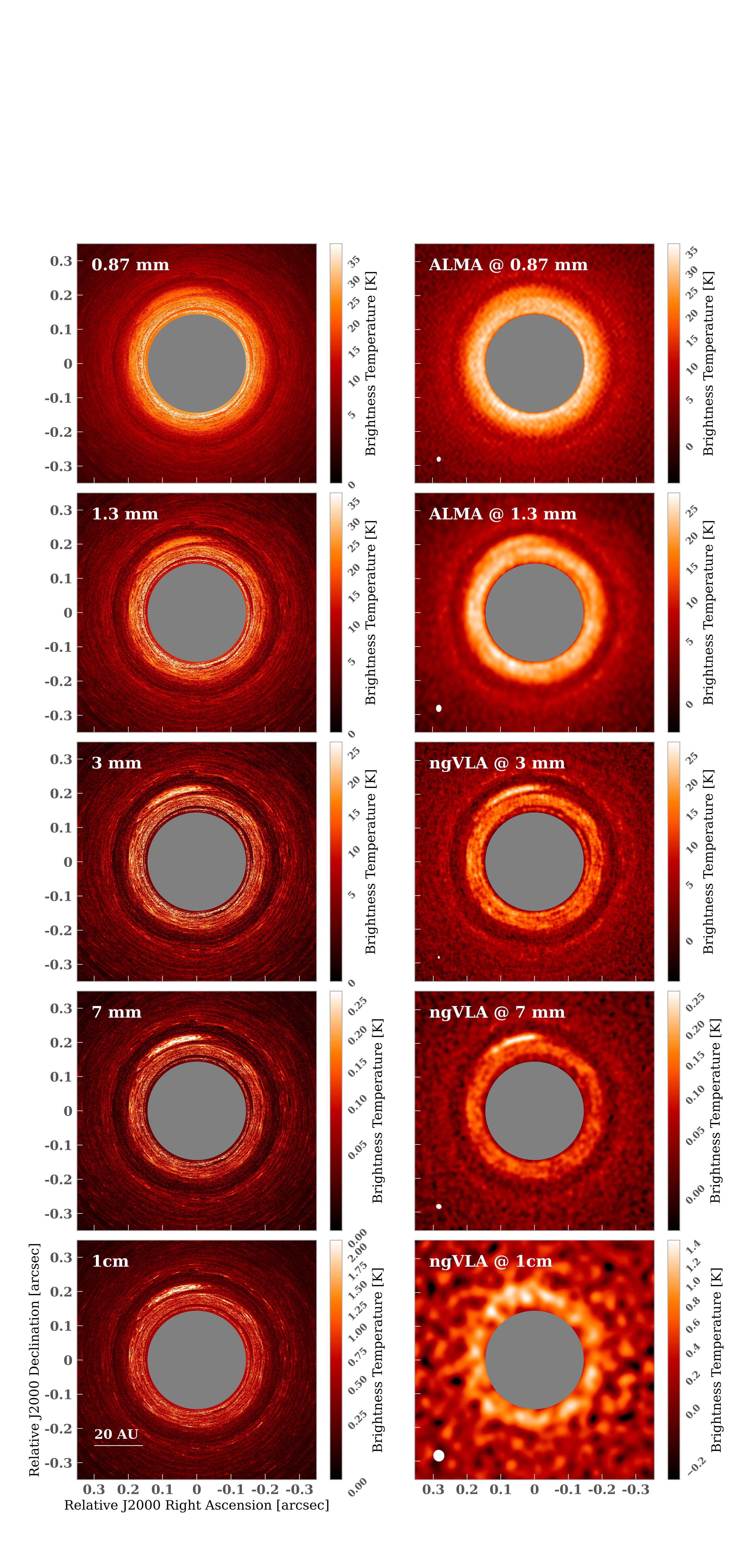}
\caption{\textit{Left:} Synthetic model images for the face-on disk for the dust thermal emission at different wavelengths, as indicated on each panel. The central circular gray cavity is the region outside of the domain of our disk simulations. Note that the color scales for the different maps are not the same, as indicated on the right of each panel. A physical scale of 20 au at the distance of the disk, i.e. 140 pc, is shown as a white horizontal bar on the bottom left panel. \textit{Right:} ALMA and ngVLA images of the dust continuum emission for the face-on disk model at various wavelengths, as indicated. On the lower left corner of each panel, the white ellipse shows the size of the synthesized beam. Beam sizes and RMS noise levels are reported in Table~\ref{table:obs}.} 

\label{fig:all_faceon}
\end{figure*}

We first recalculated the disk temperature with RADMC-3D and then extracted synthetic images for the dust thermal emission at different wavelengths between 0.87 mm and 1 cm, chosen to cover the most sensitive spectral bands to dust emission with ALMA and ngVLA (see Table~\ref{table:obs}).  We assumed isotropic scattering and the number of photons was fixed to $10^{9}$. 
The synthetic images were obtained for three disk inclinations on the sky: face-on, edge-on and 45 degrees (see left column in Figures~\ref{fig:all_faceon},~\ref{fig:all_45_deg},~\ref{fig:all_90_deg}). Each image is made of 1000 by 1000 pixels, with square pixels with angular side of 2 mas.
The resulting images are then used as input for the simulations of the ALMA and ngVLA observations. 

\begin{table*}
\centering
\begin{tabular}{ccccccc} 
\hline
\hline
 $\lambda$ [mm]  & \multicolumn{6}{c}{Dust opacity coefficients [cm$^2$ g$^{-1}$]} \\ \cline{2-7}
 & \multicolumn{2}{c}{Bin 1 ($0.1 - 10~\mu$m)} & \multicolumn{2}{c}{Bin 2 ($100~\mu$m)} & \multicolumn{2}{c}{Bin 3 ($1$~mm)} \\
 \cmidrule(lr){2-3} \cmidrule(lr){4-5}
 \cmidrule(lr){6-7} 
 & $\kappa_{\rm{abs}}$ & $\kappa_{\rm{sca}}$ & $\kappa_{\rm{abs}}$ & $\kappa_{\rm{sca}}$ & $\kappa_{\rm{abs}}$ & $\kappa_{\rm{sca}}$ \\
 \hline
 0.87 & 4.80 & 4.09 & 10.3 & 35.4 & 7.79 & 11.2\\
 1.3 & 2.15 & 0.74 & 3.04 & 6.07 & 7.22 & 12.1\\
 3.0 & 0.61 & 0.037 & 0.66 & 0.29 & 4.92 & 7.99\\
 7.0 & 0.13 & 0.00082 & 0.14 & 0.0063 & 0.76 & 9.82\\
 10.0 & 0.078 & 0.00020 & 0.078 & 0.0015 & 0.16 & 1.84\\
 \hline
 \hline
\end{tabular}
\caption{Values of the absorption and scattering dust opacity coefficients for the three grain size bins and at the wavelengths adopted for the model images presented in this study.}
\label{table:opacities}
\end{table*}

\section{Simulations of disk observations}
\label{sec:obs}

Throughout this work, we focused on the synthetic disk images at wavelengths of 0.87 mm, 1.3 mm, 3.0 mm, 7.0 mm, and 1.0 cm.
The synthetic maps for the dust continuum emission at these spectral bands were used to simulate the results of observations with ALMA and the ngVLA, following the methods presented in ~\citet{Ricci:2018}. 
For the purpose of our simulated observations, the disk was assumed to be at a distance of 140 pc, similar to the distance of the nearby Taurus and Ophiuchus star forming regions. ALMA and the future ngVLA are located in opposite hemispheres, and thus to ensure a reasonable comparison between the two telescopes, the declination on the sky of the disk was adjusted to $-24.0$ degrees for ALMA, and $+24.0$ degrees for the ngVLA simulations. 

Using the Common Astronomy Software Applications (CASA) package, we applied the \texttt{SIMOBSERVE} simulation task to transform the computed images into interferometric visibility datasets in the ($u, v$) plane. Upon simulating the ALMA observations, we used the \texttt{alma.out28} array configuration available in CASA, which includes the ALMA longest 16 km baselines, and therefore provides the best possible angular resolution at a given wavelength. We found out that this array configuration filtered out a significant fraction of the disk emission on the largest scales at 0.87 mm. To recover the filtered out emission, upon simulating the ALMA observations at 0.87 mm, we added the resultant interferometric visibilities from an additional 30 minute observation with the more compact ALMA configuration \texttt{alma.out3}
\footnote{Although this relatively short integration time with this compact array configuration allows us to retrieve the dust emission from the largest scales in the disk model discussed here, we note that actual ALMA observations would require a longer integration for optimal combination with the observations with a more extended configuration.}. Upon simulating the ngVLA observations, we used the ngVLA Main Array Configuration Rev. B, which is composed of 214 antennas of 18 meter diameter with baselines ranging up to $\sim$~1000 km ~\citep{Carilli:2018, CarilliandErickson:2018, SelinaandMurphy:2017}. We did not include the 30 additional stations in the ngVLA Long Baseline Array, which correspond to baselines up to $\sim$~9000 km, because of the poor sensitivity to the emission from our sources of interest ~\citep{Carilli:2018}.

\begin{table*}
\centering
\begin{tabular}{ccccc} 
\hline
\hline
Telescope & $\lambda$ & Flux Density & Resolution Beam & RMS Noise \\
 & [mm] & [mJy] & [mas $\times$ mas] & [$\mu$Jy/beam] \\
\hline
      \multirow{3}{*}{\textbf{ALMA}} 
      & 0.87 & 450 & 13 $\times$ 10 & 12.0 \\ 
       & 1.3 & 150 & 19 $\times$ 15 & 6.0 \\ 
       & 3 & 16 & 44 $\times$ 34 & 3.9\\ 
        \hline 
      \multirow{3}{*}{\textbf{ngVLA}} 
       & 3 & 16 & 8 $\times$ 6 & 0.34  \\ 
       & 7 &  0.7 & 14 $\times$ 11 & 0.085 \\ 
       & 10 & 0.1 & 34 $\times$ 30 & 0.10 \\ 

 \hline
 \hline
\end{tabular}
\caption{List of the main observational characteristics of the simulated ALMA and ngVLA observations for our disk model with face-on inclination, as described in Section~\ref{sec:obs}. The third column represents the flux densities derived by integrating the surface brightness over the whole model disk.}
\label{table:obs}
\end{table*}




In the ALMA case, the \texttt{SIMOBSERVE} task automatically assigns a noise level to the sampled visibility points. 
With the exception of the ALMA observations at 0.87 mm which also include a short integration with a more compact array configuration, the visibility datapoints sampled by the simulated ALMA observations correspond to an aperture synthesis spanning 10 hours centered on transit. 
In the ngVLA case, we instead adopted the procedure detailed in ~\citet{Carilli:2016} and utilized \texttt{SETNOISE} in the \texttt{SIMOBSERVE} task to corrupt the visibilities and allocate a set noise level.
According to the nominal parameters listed in the ngVLA reference design \citep{Selina:2018}, the rms noise values can be obtained by the ngVLA with an integration of about 10 hrs.
Synthetic observations with longer integration times for both ALMA and ngVLA are presented in Section~\ref{sec:results}. 

The obtained visibilities were then imaged using the \texttt{CLEAN} algorithm in the CASA package.
The imaging quality depends on a balance between the array sensitivity and synthesized beam ~\citep{Carilli:2018}, which can be tailored using different values of the robust parameter $\rm{R}$ for the Briggs weighting scheme, as well as by applying a tapering of the interferometric visibilities on the $(u,v)$ plane. 
For the ngVLA maps we found a good compromise between angular resolution and sensitivity using $\rm{R} = -1$, whereas the ALMA images were produced with $\rm{R} = -2$ \citep{Ricci:2018}. Additionally, in order to improve the signal-to-noise ratio at the expense of angular resolution in the ngVLA maps, we tested several outer tapers to the visibilities on the longest baselines, correspondent to angular scales between 5 mas and 30 mas.

The main observational properties including resolution beam size and rms noise for the ALMA and ngVLA maps presented in this work are summarized in Table~\ref{table:obs}. 

\section{Results of simulated observations} 
\label{sec:results}

In this section, we outline the results of the disk simulations presented in Section~\ref{sec:sims} with a focus on the simulated observations with ALMA and the ngVLA. We divide our discussion into three different disk orientations in the sky, i.e. face-on, intermediate inclination of 45 degrees, and edge-on, which allows us to highlight different substructures in the disk.

\subsection{Simulated ALMA and ngVLA observations for a face-on disk}
\label{sec:simulated_obs}

We start by discussing the results for a disk with face-on orientation in the sky. This orientation allows us to best probe substructures both in the radial and azimuthal directions. 
Figure~\ref{fig:all_faceon} shows the synthetic model images and ALMA and ngVLA maps for the dust continuum emission at wavelengths between 0.87 mm and 1 cm. 

The most prominent substructure which is clearly identified in all the maps of Figure~\ref{fig:all_faceon} is the circular gap at about 0.2 arcsec from the star, corresponding to $\sim$~34 au at the distance of the disk. The dark concentric gap is visible at all wavelengths, with the highest contrast from neighboring dust emission obtained at 3 mm. Moreover, the gap width and depth vary with wavelength. The model predicts a radial widening of the gap at longer wavelengths as a consequence of the larger value of the Stokes parameter of the larger grains which dominate the emission at longer wavelengths, as well as of the lower optical depth of their emission probed at lower frequencies. The wavelength-dependent radial profiles of the surface brightness of our disk model are presented in Appendix~\ref{sec:appendix_radial_profiles}.

Towards the north side of the ring on the inward side of the gap, the dust emission shows an azimuthally asymmetric increase, or a \textit{bump}, in the local surface brightness. This local accumulation of dust is similarly visible in all the maps and its morphology varies with wavelength. At 0.87 mm, the local dust emission is difficult to distinguish from neighboring emission and appears radially larger and azimuthally wider. At longer wavelengths, the contrast of the dust bump is higher and corresponds to the brightest region within the radial domain probed by our simulation. Furthermore, the emission from this bump structure shrinks both radially and azimuthally at longer wavelengths. This is a consequence of the spatial segregation of dust grains with different sizes within a gaseous vortex (see Section~\ref{sec:discussion}). The wavelength-dependent azimuthal profiles of the surface brightness in this disk region are presented in Appendix~\ref{sec:appendix_azimuthal_profiles}.

Another interesting set of substructures which is present in the model synthetic maps is a collection of radially narrow circular arcs beyond the main gap discussed above, i.e. at stellocentric radii $r > 40-45$ au. Similar structures are seen also inward of the gap, all the way to the inner boundary of the radial domain of the simulation. In fact, the region of the inner edge of the gap appears to be made of a \textit{bundle} of narrow arc-like substructures.  
Like for the other substructures discussed above, also these ones are less visible at higher frequencies because of a combination of less efficient dust trapping of smaller grains and higher optical depths.

The right column in Figure~\ref{fig:all_faceon} showcases the results of our ALMA and ngVLA simulated observations. 
From the maps shown in the figure, it is clear that the ngVLA at 3 mm offers the best 
opportunity to resolve and characterize the structures presented in this section.  
At 3 mm, the ngVLA provides the best resolution, as well as better sensitivity to dust emission than at longer wavelengths.
In the following we discuss the potential of ALMA and ngVLA to characterize the substructures discussed here, and provide some information on their physical origin.

\begin{figure*}[t!]
\centering
\includegraphics[height=23cm, trim = {1.5cm 2cm 1.5cm 12cm}]{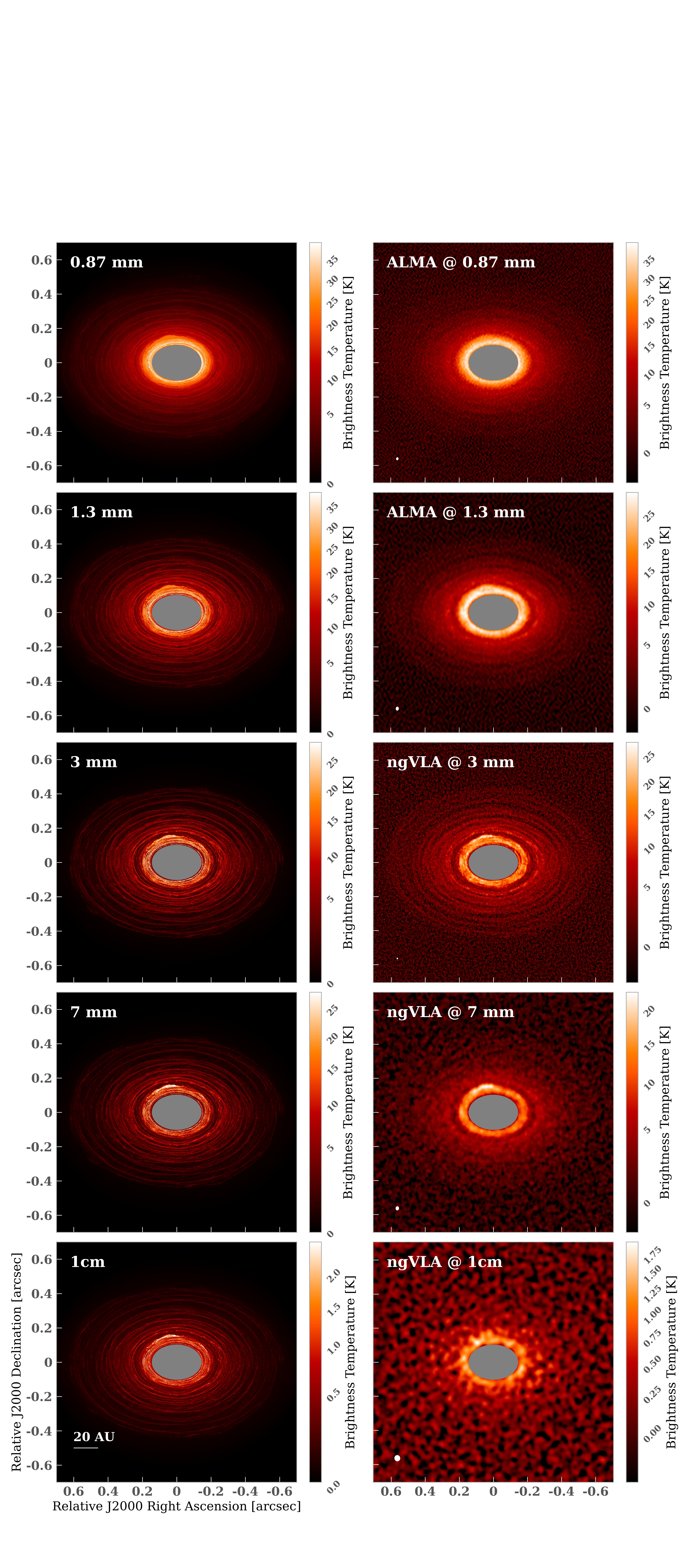}
\caption{Same model as Figure~\ref{fig:all_faceon} but viewed 45$^{\circ}$ from face-on.}
\label{fig:all_45_deg}
\end{figure*}

\subsubsection{Inner gap at 30 au from the star}

All the ALMA and ngVLA maps reveal the presence of a prominent gap at stellocentric radius $r \approx$~30 au surrounded by annular regions of relatively high brightness. At first glance, a comparison between the different maps and the corresponding model images highlight the potential of the ngVLA to capture the width and contrast of this inner gap and surrounding dust concentrations. This is especially true at 3 mm, in which the radial width of the gap is covered by $4-5$ resolution beams.  
In the ALMA maps at 0.87 mm and 1.3 mm, the gap is significantly narrower and with lower contrast. At 3 mm, the size of the ALMA synthesized beam approaches the radial width of the gap, resulting in a poorly resolved substructure. For this reason, we do not show this map in this paper. 

These azimuthally symmetric substructures are similar to the concentric gaps and rings which are the most frequent structures observed so far in protoplanetary disks  \citep{Andrews:2018,Huang:2018,Cieza:2020}. 
The presence of these  substructures in the dust emission is likely originated from similar structures, although with lower contrast, in the underlying gas distribution: the overdense regions in the gas component correspond to radial pressure maxima, and the inward radial drift of small particles can be significantly slowed down or even halted, hence creating a strong enhancement in the density and emission of dust~\citep[e.g.,][]{Pinilla:2012}. In young disks, with or without pre-existing planets, these regions of accumulation of large grains can trigger the formation of planetesimals \citep[e.g., via streaming instabilities,][]{Carrera:2020}, the building blocks of planets. Our models of a VSI-dominated disk predict the formation of a dark gap located at $\sim$~30 au from the central star, with an accumulation of dust at the edges of the gap.
This is due to a local decrease of the accretion stress produced by the VSI at about 30 au from the star, where thermal relaxation is slower than the orbital period and approaches the rate below which VSI is cut off \citep{Flock:2020}.

\begin{figure*}[t!]
\centering
\includegraphics[height=20cm, trim = {0cm 1cm 1.5cm 5cm}]{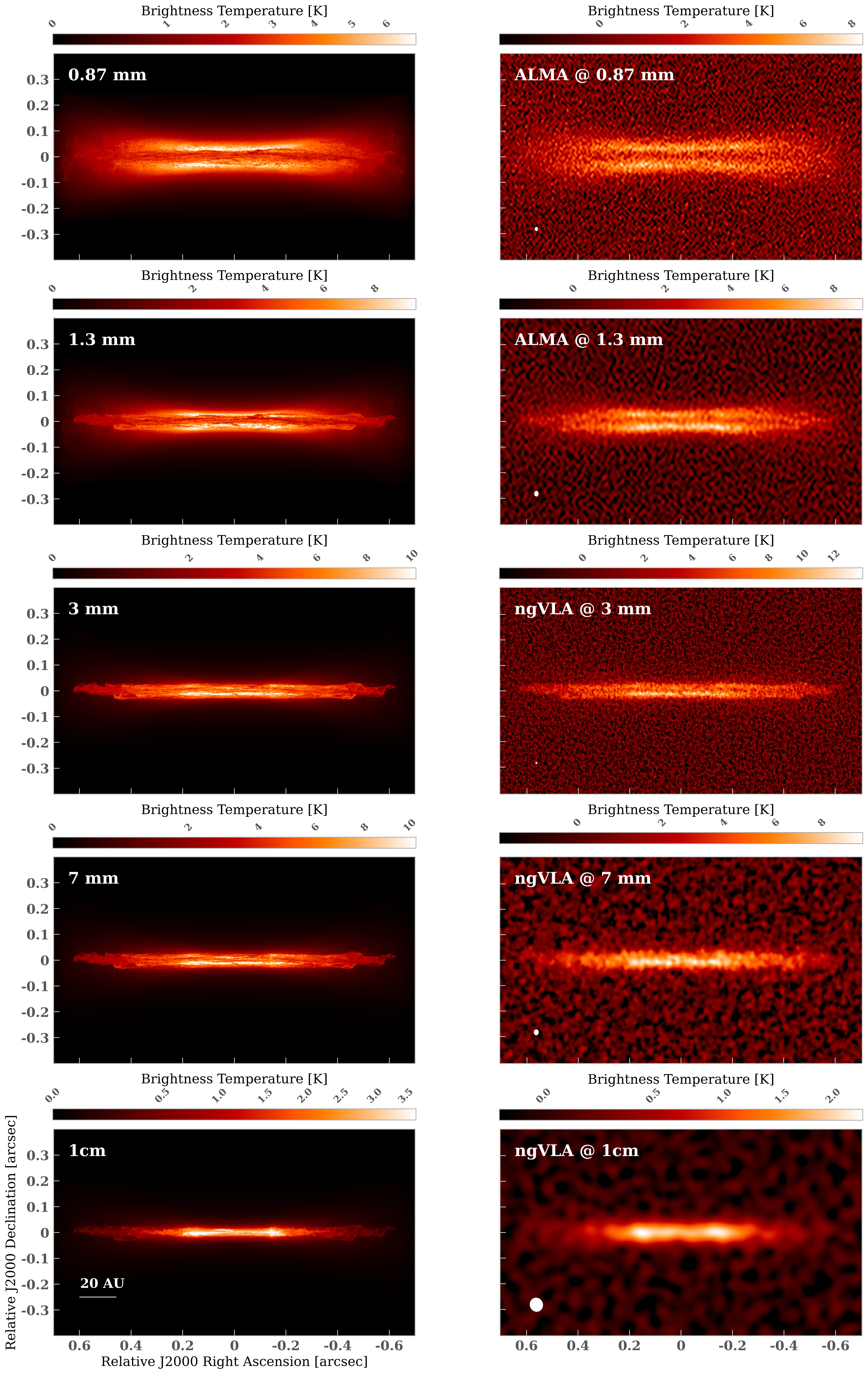}
\caption{Same model as Figure~\ref{fig:all_faceon} but viewed edge-on.}
\label{fig:all_90_deg}
\end{figure*}

\subsubsection{Dust Concentration towards the Vortex Region}

As described in Section~\ref{sec:simulated_obs}, towards the north side of the ring on the inward side of the gap, the dust emission shows an azimuthal increase, or a \textit{bump}, which is due to dust concentration inside a gaseous vortex (see Section~\ref{sec:discussion}).
As shown in Figure~\ref{fig:all_faceon}, this region of dust concentration is not well resolved by ALMA along the radial direction. Due to the limited angular resolution, in the ALMA maps the dust bump appears attached to, and difficult to distinguish from the bright, inner, ring-like structure. Conversely, the bump is well resolved and well-separated from the inner disk emission with the ngVLA at 3 mm and 7 mm. Hence, the ngVLA would be capable of resolving the local concentration of dust at $\sim$~1 au resolution, for a disk at a distance of 140 pc.

This is reminiscent of another type of substructures found in high-angular resolution observations of young disks, associated with strong azimuthal asymmetry. Examples of these structures have crescent-shaped or bump features, as those found in systems like Oph IRS 48 \citep{vandermarel:2013}, HD 142527 \citep{Casassus:2015},  MWC 758 \citep{Boehler:2018}, and HD 143006 \citep{perez2018}, among others. Our disk model predicts the presence of a high-contrast bump located on the inner edge of the dark gap, although not as prominent as the crescent-shaped structures seen in disks like IRS 48 or HD 142527. The feature in our model is due to dust concentration inside a vortex generated in the gas component of the disk simulation.
While VSI is known to hamper grain growth due to the induced turbulent velocities and resultant large vertical scale height of dust~\citep{flock20}, these unfavorable consequences of VSI to planet formation can be reduced in some disk regions through the formation of local pressure maxima that trap particles, reduce inward radial drift and enhance particle growth~\citep{lin2015}. The formation and survival of gaseous vortices in disks subject to VSI have been investigated via 3D global simulations of gaseous disks by \citet{Manger:2018}. 
This study has shown that the VSI is capable of forming relatively large and long-lived vortices ($\sim$ hundreds of orbits).

Besides confirming this general result, our simulations are able to follow the motion and local concentration of dust particles with different sizes within the vortex. The dust motions in the vortex result in more efficient concentration for larger grains, and this reflects into a more compact bump along the azimuthal coordinate at longer wavelengths (see also Appendix~\ref{sec:appendix_azimuthal_profiles}). Another factor that increases the azimuthal extent of the bump at wavelengths of 0.87 and 1.3 mm relative to longer wavelengths is the inverse relation between dust optical depth and wavelength. 

\subsubsection{Arc-like substructures}
\label{sec:arcs}

As discussed in Section~\ref{sec:simulated_obs}, the model synthetic images show that the dust emission both outside of the gap as well as at lower stellocentric radii is made of radially narrow arc-like substructures. 
These features are more prominent at longer wavelengths, and Figure~\ref{fig:all_faceon} shows that the brightest among these structures can be detected with the ngVLA at 3 mm.
Although the detection of these narrow features is rather marginal for a face-on disk, especially in the outer disk regions, in the next section we show that similar observations for an inclined disk would allow to better detect and characterize these substructures.  

\begin{figure*}[thb!]
\centering
\includegraphics[width=\linewidth]{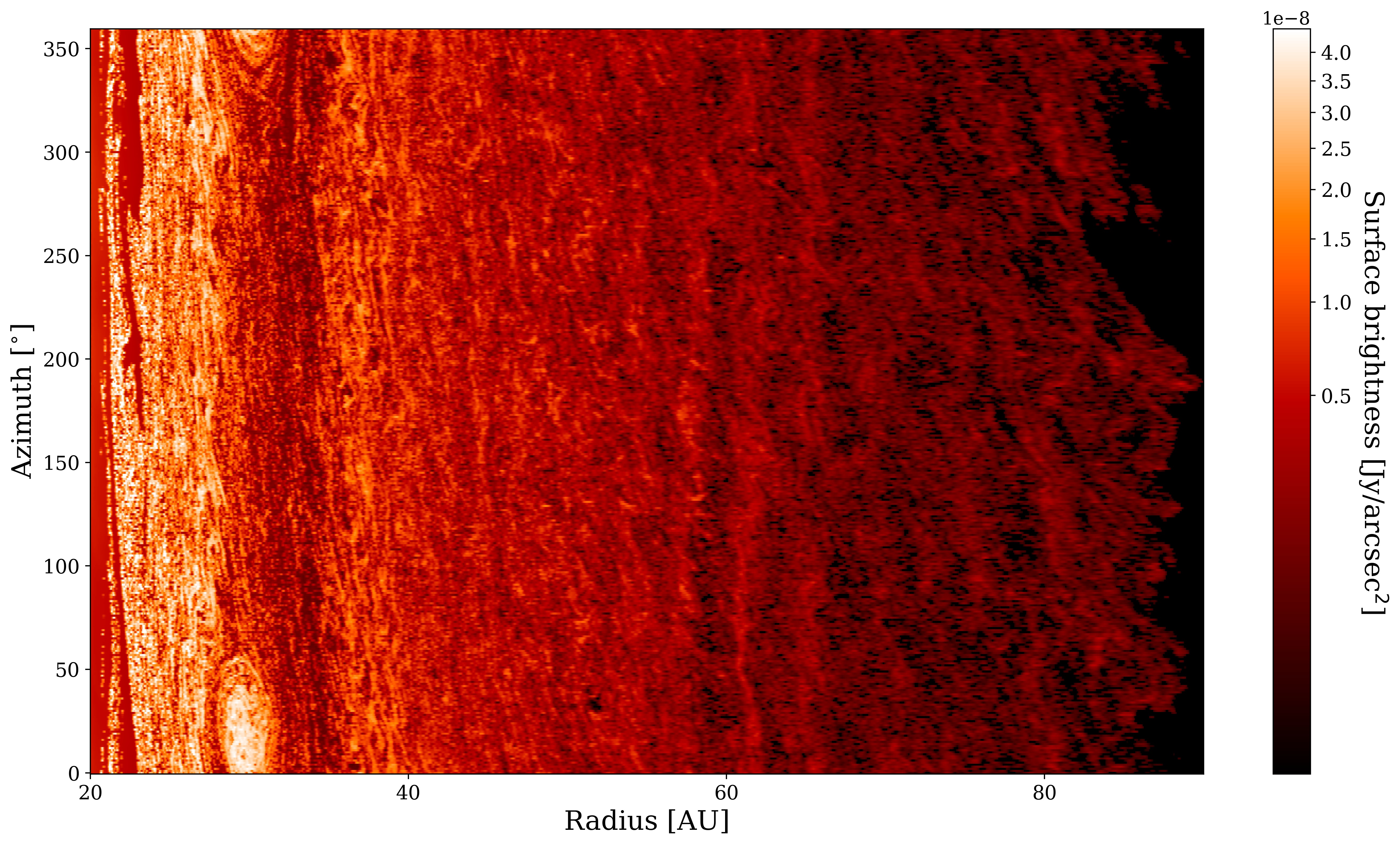}
\caption{Projection onto polar coordinates of the dust continuum emission for the disk model at 3 mm.}
\label{fig:polar_3mm}
\end{figure*}

These dust substructures are due to local changes of the grains radial drift caused by the turbulence and the mostly axisymmetric gas motions from the vertical shear instability. As shown in Figure~\ref{fig:polar_3mm}, these features are not complete rings, even though they can have large azimuthal extents. 
Arc-like substructures have been detected with ALMA in the dust continuum emission of some young disks \citep[e.g., ][]{vandermarel:2016,Cazzoletti:2018,Isella:2018}. However, the detected structures have much higher contrasts and are much larger than those predicted in this work, which have radial widths of $\sim 1-2$ au in the outer disk, and are even narrower at stellocentric radii $r < 30$ au.

To our knowledge, features with these characteristics have not been predicted by any other physical mechanism proposed in the literature, and their detection in future high angular resolution observations could provide evidence for VSI acting in real systems.

\subsection{ALMA and the ngVLA simulated observations of a disk with inclination of 45 degrees}

As shown in Figure~\ref{fig:all_45_deg}, the synthetic images of the same disk model but with an inclination in the sky of 45 degrees better expose the emission from the set of azimuthal arc-like features, especially in the outer disk regions. The higher contrast provided by the inclined disk is a consequence of the relationship between the inclination angle $i$ and the line-of-sight optical depth ($\tau_{\nu}^{i}$) of the emission: 
\begin{equation}
    \tau_{\nu}^{i}  = \frac{\Sigma(r)\kappa_{\nu}^{i}}{\cos{i}}; 
\end{equation}
where $\Sigma(r)$ is the dust surface density at stellocentric radius $r$ and $\kappa_{\nu}$ is the dust opacity coefficient. The increased inclination of the disk additionally reduces the visibility of the dust bump, which, because of the projection on the plane of the sky, appears closer to the inner edge of the gap. 

The results of the simulated observations with ALMA and the ngVLA are displayed in the right column of Figure ~\ref{fig:all_45_deg}.
ALMA starts to detect the brightest arcs in the disk outskirts at 1.3 mm, whereas the ngVLA map at 3 mm better unveils and separates the emission from more arcs in those regions.
Due to the lower sensitivity of the ngVLA to the dust emission at longer wavelengths, these structures are not visible at neither 7 mm nor 1 cm.
In Section~\ref{sec:discussion} we present the results of similar observations but with significantly longer integration times, and therefore higher signal-to-noise ratios on the maps.

\subsection{ALMA and the ngVLA simulated observations for an edge-on disk}

The left column in Figure ~\ref{fig:all_90_deg} displays the predicted dust emission for a perfectly edge-on disk, i.e. $i$ = 90$^{\circ}$. The model images display a series of interesting features. At the shortest wavelength, 0.87 mm, the disk exhibits a prominent dual-lobe structure. The disk midplane is dark and sharply contrasts against the brighter emission from vertical layers above and below the disk midplane. At longer wavelengths, i.e., at 1.3, 3, and 7 mm, the dark lane at the midplane is still visible, but with a much lower vertical extent as the dust emission sandwiching the dark midplane flattens at longer wavelengths. At 1 cm, the dark lane is no longer distinguishable. 

The ALMA and ngVLA simulated maps for this edge-on disk are presented in the right column in Figure ~\ref{fig:all_90_deg}. In the ALMA map at 0.87 mm, the two vertical lobes above and below the midplane are well separated and the darker midplane is also visible. Similarly, the ALMA observations at 1.3 mm also capture this sandwich-like structure, although with lower contrast. A comparison between these predictions and recent results of ALMA observations of edge-on disks is presented in Section~\ref{sec:DSHARP}. 

As for the ngVLA, the observations at 3 mm resolve the vertical height of the disk despite the significantly thinner extent than at shorter wavelengths, and the less bright wave-like substructures at the opposite ends of the disk are also detected. These structures are not detected by the ngVLA at 7 mm and 1 cm, and the dark midplane is no longer visible either. 
At 1 cm, the ngVLA resolution does not allow to spatially resolve the vertical scale height, which causes the disk to appear vertically more extended than at 3 and 7 mm. These observations can also detect features which are asymmetric about the midplane. These are generally due to the vertical motions of dust grains induced by the VSI. An investigation of the time variation of these features will be presented in a future paper  (Carrasco et al., in prep.).


\section{Discussion} 
\label{sec:discussion}

Our work was motivated by the demand of physical disk models that may explain the recent results of high angular resolution observations of young disks, especially for the dust continuum emission at sub-mm/mm wavelengths. 
In particular, the largest high-resolution ALMA survey of disks conducted so far in nearby star forming regions has unveiled the presence of substructures in the thermal dust emission of all the observed disks \citep[\textit{DSHARP},][]{Andrews:2018}. These substructures have various morphological features, such as dark gaps, rings and arcs, inner cavities, spiral arms, and local crescent-shaped substructures.

To date, the majority of theoretical studies on the origin of these features have focused on the disk dynamics in the presence of one or more embedded planets. Although the disk-planet interaction remains a viable hypothesis to explain several of the observed substructures, this interpretation may not be unique, especially as long as the presence of these planets cannot be ascertained observationally.

Moreover, models of planetesimal formation predict that small solids with sizes lower than $\approx 1-10$ cm have to accumulate locally, or \textit{trapped}, in some regions of the disk, before they acquire very high speeds due to the aerodynamical interaction with the gas \citep{Weidenschilling:1977,Brauer:2007}.
For this trapping of small solids to occur, relatively stable regions of local density/pressure maxima have to form, and the most plausible triggering mechanisms are hydrodynamical instabilities, with or without the effect of a magnetic field threading the disk.   

The results presented in Sections~\ref{sec:obs} and \ref{sec:results} indicate that the vertical shear instability can trigger the local trapping of $\sim$ mm-sized particles, and also that several dusty substructures predicted from our model can be detected and characterized with ALMA and a future ngVLA at sub-mm/mm wavelengths.

In this section, we discuss some important characteristics of the substructures predicted by our model and presented in Section~\ref{sec:results}, and we then compare the theoretical predictions for this model with the results of current and future ALMA and ngVLA observations.

{\subsection{Lifetime and other characteristics of the disk substructures}

We have discussed the appearance of
axisymmetric and non-axisymmetric structures in a VSI-turbulent
protoplanetary disk.  The axisymmetric rings and gaps result from
radial variations in the accretion stress and so are likely to
survive for many orbital periods.  The most prominent rings and gap are consequences of the decline with radius in the thermal relaxation timescale as a fraction of the orbital period. Beyond about 30~au, thermal relaxation takes less than 1/10th of an orbit
\citep[see][]{Flock:2020}.  Strong VSI then produces a strong accretion
stress, which over the radial flow timescale carves the
perturbation we see in the gas surface density.

We note that the radial location of the gap critically depends on the total amount of small grains in the disk. Those grains define the dust opacities which control the cooling and heating of the disk. For example, in a massive disk with larger amounts of small grains than considered in the disk model presented here, we would expect the gap to extend at stellocentric radii beyond 30 au. On the other hand, if small grains are less abundant, e.g. because of the possible combined effects of dust vertical settling towards the disk midplane and grain growth, then we would expect the gap at stellocentric radii lower than 30 au. 

Contrary to this large scale axisymmetric structure, the non-axisymmetric substructures discussed in Section~\ref{sec:arcs} show time variations on the local orbital timescale. These variations are associated to small scale dust clumping by tiny vortices \citep[see Appendix in][]{Flock:2020}, and also radial dust concentration which occurs in between the large scale vertical upward and downward VSI motions \citep{Flores2020}. These two features are short lived, with variations on timescales of several orbits \citep[but then similar other features reappear periodically as they are generated by small vortices which appear frequently during the course of the simulation described in][]{Flock:2020}. However we expect that, given the large stellocentric distances of these substructures, not even observations separated by decades would be able to detect their time variation. We plan on investigating the time variation of the structures predicted by our disk model in a future work (Carrasco et al. in prep.).}

\begin{figure}
    \centering
    \includegraphics[width=\columnwidth]{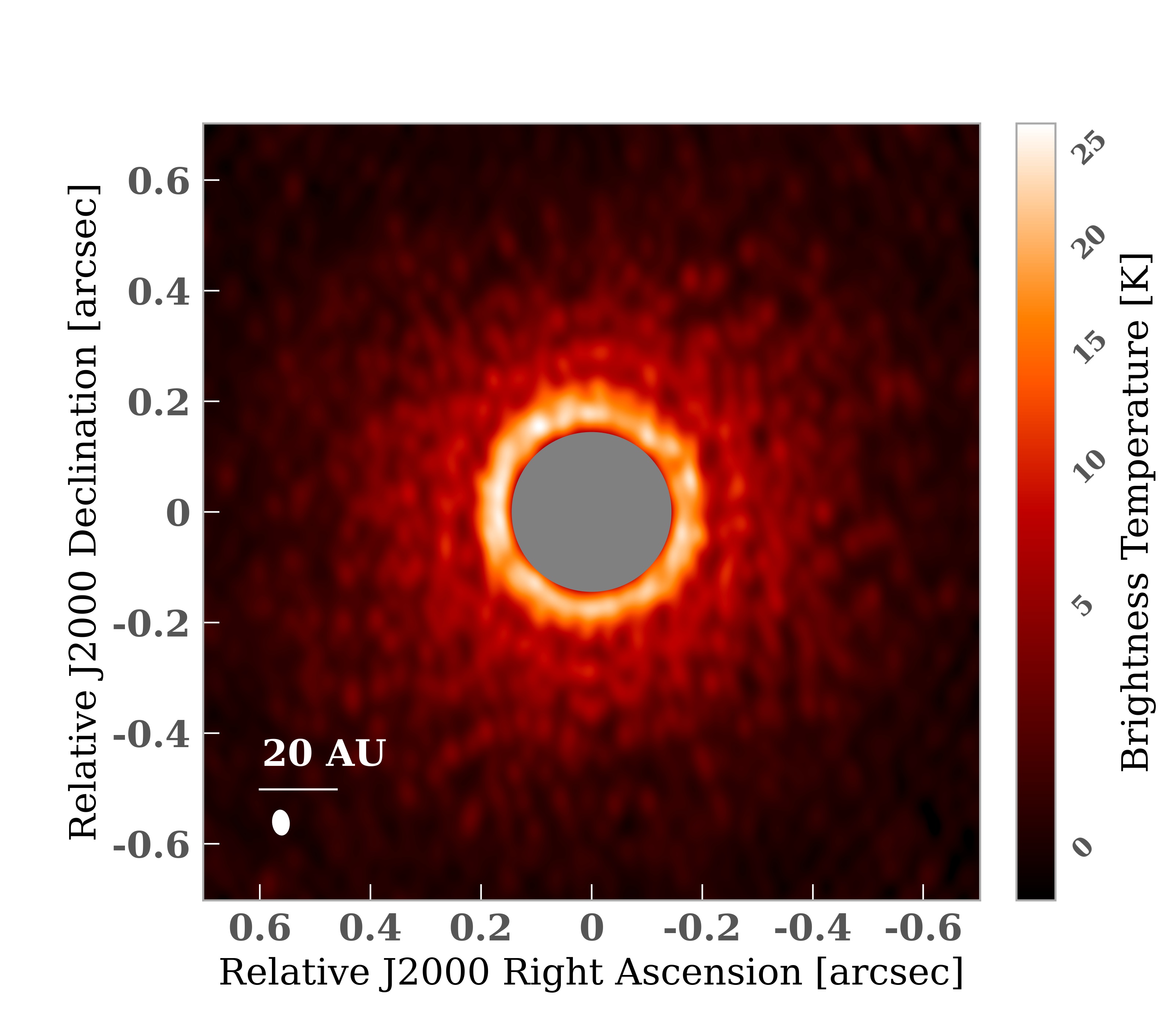}
    \caption{Image of the ALMA dust continuum emission at 1.3 mm obtained with observational characteristics that resemble those obtained for the ALMA DSHARP project, as detailed in Section~\ref{sec:DSHARP}. Imaging was performed with a Briggs robust parameter of 0. The resultant RMS noise on the map and synthesized beam, shown in the lower left corner, are about 15.0 $\mu$Jy/beam and 43 mas $\times$ 28 mas, respectively.}
    \label{fig:dsharp_esque_map}
\end{figure}

\begin{figure}
    \centering
    \includegraphics[width = \columnwidth]{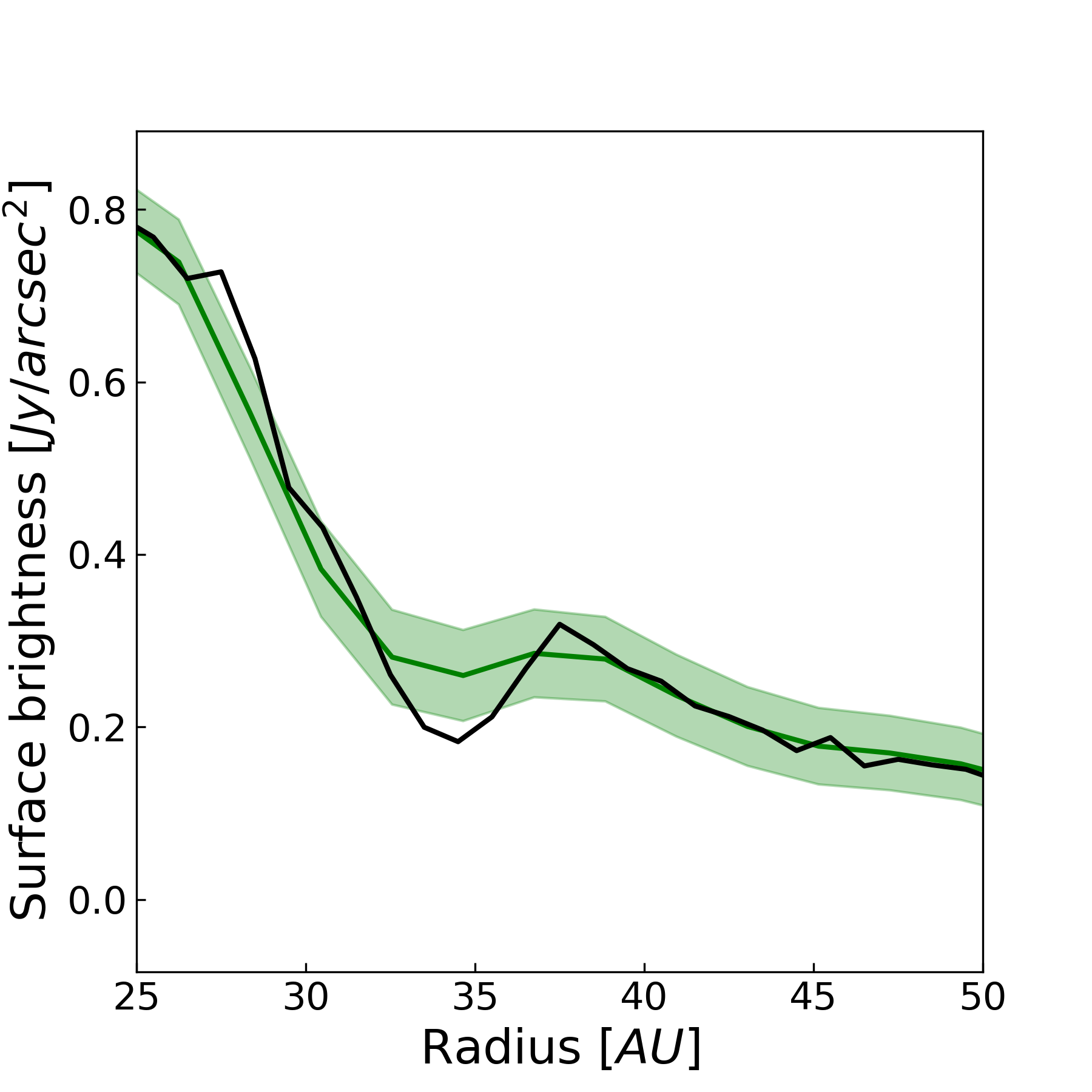}
    \caption{Radial profile for the dust surface brightness at 1.3 mm from the synthetic ALMA map shown in Figure~\ref{fig:dsharp_esque_map}, after averaging over the azimuthal coordinate (green line). The shadowed area in green represent the 1$\sigma$ confidence intervals at each radius, which are equal to the ratio between the standard deviation of the surface brightness in each annulus and the square root of the number of resolution beams within the annulus area. The black line is the radial profile obtained from the synthetic model image before the simulated ALMA observations.}
    \label{fig:dsharp_esque_rad_prof_50au}
\end{figure}

\begin{figure*}
    \centering
    \includegraphics[width = \linewidth]{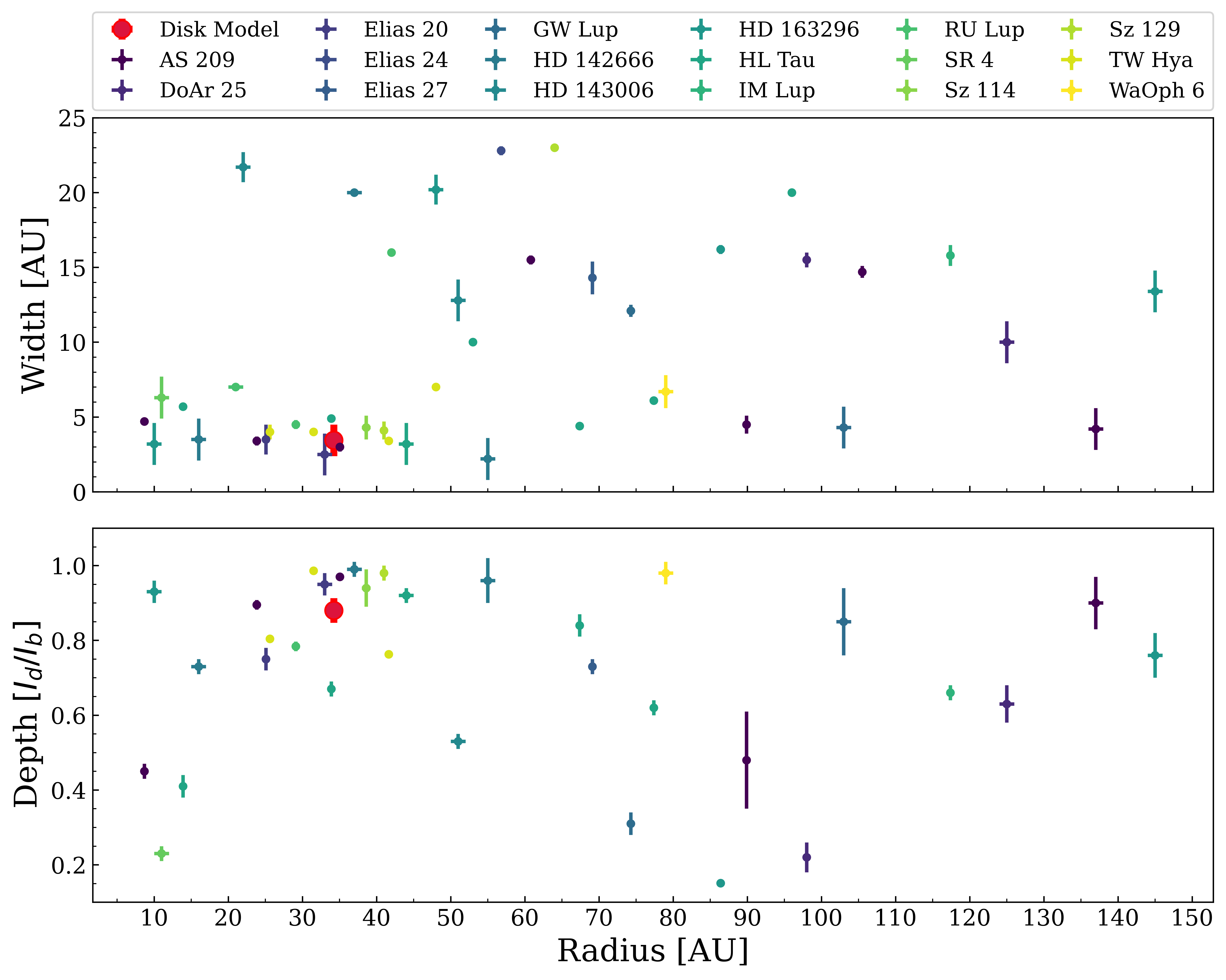}
    \caption{\textit{Top:} Plot of the radial locations and widths of the dark gaps in HL Tau, TW Hya, and a sample of disks from the DSHARP survey at a wavelength of about 1.3 mm. Uncertainties are marked by error bars. The properties of the dark gap predicted by the disk model at 1.3 mm presented in this work are indicated by the red filled circle. \textit{Bottom:} Same as above but for the dark gap depths.}
    \label{fig:rad_dep_wid}
\end{figure*}

\subsection{Comparison with recent observations of analogous features with ALMA}
\label{sec:DSHARP}

The results presented so far show that models of disks undergoing VSI predict the existence of various substructures in the dust thermal emission.  
In order to see whether observations of the dust continuum from disks can test this prediction, we have to first derive images from our models which mimic the characteristics of those observations. 
To do this, we consider the observations from the DSHARP program, which has provided the largest survey conducted so far of nearby disks at high angular resolution with ALMA.
To simulate ``DSHARP-like'' observations for our disk model, we considered the synthetic image at 1.3 mm (Figure~\ref{fig:all_faceon}), and ran simulations for ALMA observations as discussed in Section~\ref{sec:simulated_obs}, but with the less extended ALMA configuration \texttt{alma.out24} and shorter integration time. By deconvolving the interferometric \textit{dirty image} using Briggs weighting with a robust parameter of 0.0, the resulting image has an RMS noise level of 15.0 $\mu$Jy/beam and a synthesized beam with a size of 43 mas $\times$ 28 mas, which are similar to the typical observational parameters of the DSHARP program \citep{Andrews:2018}.

This image, which is presented in Figure~\ref{fig:dsharp_esque_map}, together with the radial profile of the surface brightness in Figure~\ref{fig:dsharp_esque_rad_prof_50au}, indicate that the gap structure at $\sim$ 34 au from the star is not well resolved (spatial resolution of about 6 au $\times$ 4 au), and appears as a ``plateau''-like feature in the radial profile between stellocentric radii of about 32 and 39 au. Similar features have been found in a number of disks within the DSHARP survey~\citep{huang18b}. 
In particular, Figure~\ref{fig:rad_dep_wid} shows the radial positions, radial widths and depths of dark gaps detected by the DSHARP program with the addition of the HL Tau and TW Hya young disks~\citep{Huang:2018}, together with the properties of the gap predicted by our model in red. The parameters for the predicted gap were obtained following the procedure  in~\citet{Huang:2018}, whereas the uncertainties were estimated by considering the standard deviation of the parameter values after arbitrarily varying the width of the radial bins in the radial profile. The radially narrow, shallow plateau-like gap predicted by our disk model falls within a population of dark annular substructures located in the inner $45-50$ au of the observed protoplanetary disks.

Figure~\ref{fig:dsharp_esque_map} also shows that the azimuthal bump structure towards the northern part of the inner edge of the gap is not separated from the bright emission from the inner ring and cannot be identified from this map. This result suggests that a fraction of the azimuthally symmetric rings observed by the DSHARP program may actually be hiding regions of local dust concentration that only observations with better resolution and/or at longer wavelengths can unveil. 
Moreover, the arc-like substructures predicted by the model are buried in the noise in the outer disk regions, and not resolved in the inner ring close to the inner boundary of the simulation.

ALMA observations in the dust continuum at 0.85 mm of HH 212, a nearby protostellar system deeply embedded in a compact molecular cloud core in the L1630 cloud of Orion, have yielded the first detection of a dark lane in correspondence with the disk midplane at sub-mm wavelengths~\citep{Lee17}. The disk is nearly edge-on on the sky, and features two bright lobes sandwiching the dark disk equatorial plane. Physical disk models which account for dust growth and vertical settling predict the presence of millimeter-sized dust grains with higher densities and lower temperatures in the disk midplane rather than in the vertical bright layers above and below it~\citep{Leee1602935}.

\begin{figure}
\centering
\includegraphics[width=\columnwidth, trim = {0.4cm 0.6cm 1.6cm 0.7cm}]{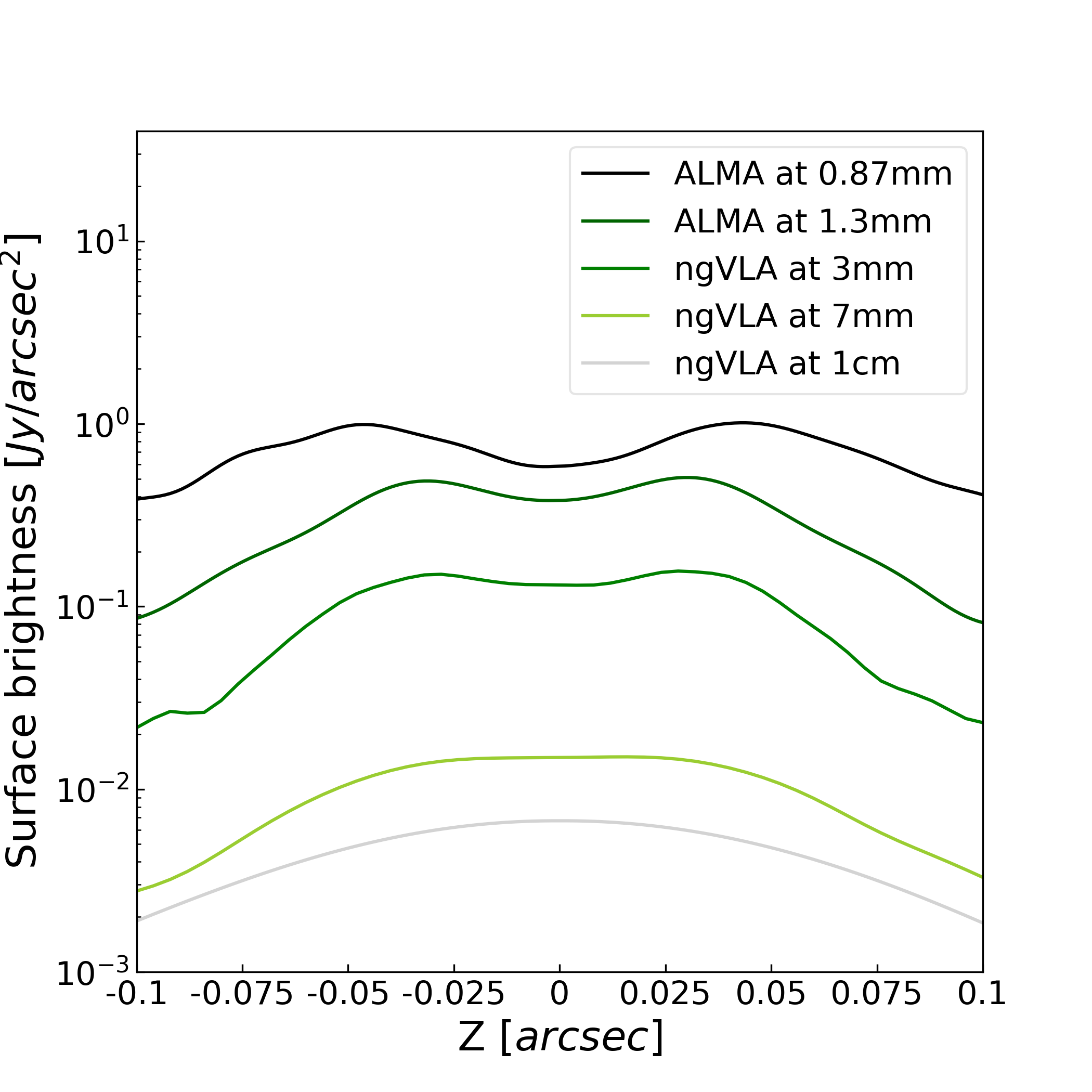}
\caption{Vertical profiles of dust surface brightness for the simulated observations of the edge-on disk at the wavelengths as labeled. The vertical profiles were extracted from a vertical cut centered at the disk center and with a horizontal width of 0.2 arcsec.}
\label{fig:vertical_profile}
\end{figure} 

Our models predict similar structures for a perfectly edge-on disk at sub-mm wavelengths, with dust vertically stirred due to VSI-driven turbulence~(Figure~\ref{fig:all_90_deg}). 
Although a detailed comparison with the ALMA observations obtained for HH 212 would require an adjustment of our model parameters to this specific case, the double-lobed, hamburger-like appearance is reminiscent of the HH 212 bright upper and lower vertical layers and dark midplane with comparable aspect ratio.
We also note that a recent ALMA survey was conducted on a sample of twelve edge-on disks in Taurus, Ophiuchus and Chamaeleon I star forming regions \citep{villenave:2020}. These observations did not find dark midplane lanes at wavelengths of 0.89, 1.33 and 2.06 mm, but the angular resolution of this survey is about 0.1$''$, i.e. a factor of $\approx 10\times$ worse than in the simulated observations presented in this work. Future observations with better resolution may unveil dark midplane lanes in nearby edge-on disks. 

Our results show that ALMA and the ngVLA would be able to resolve the vertical extent of edge-on disks up to $\approx$ 7 mm. In Figure~\ref{fig:vertical_profile} we superimpose the surface brightness of the simulated observations from both interferometers for the edge-on disk model along a vertical cut through the disk center. At shorter wavelengths the disk lobes are brighter than the midplane emission by a factor up to 40\% and the dark lane is wider, which is in line with the findings for the HH 212 system~\citep{lin20}. At the same time, the emission displays a decreasing vertical extent with increasing wavelength, a feature that results from a combination of the lower optical depth of the dust emission at longer wavelengths and more efficient vertical settling for grains with larger size. 
The synergy between high-angular resolution observations with ALMA and a future ngVLA has the potential to better probe the dependence with wavelength of the vertical height of the dust emission. This would probe the size-dependent vertical settling of grains towards the midplane and test the predictions from disk models with VSI. 

\subsection{Predictions for long integration hi-res observations with ALMA and ngVLA}
\begin{figure}[t!]
\centering
\includegraphics[width=\columnwidth]{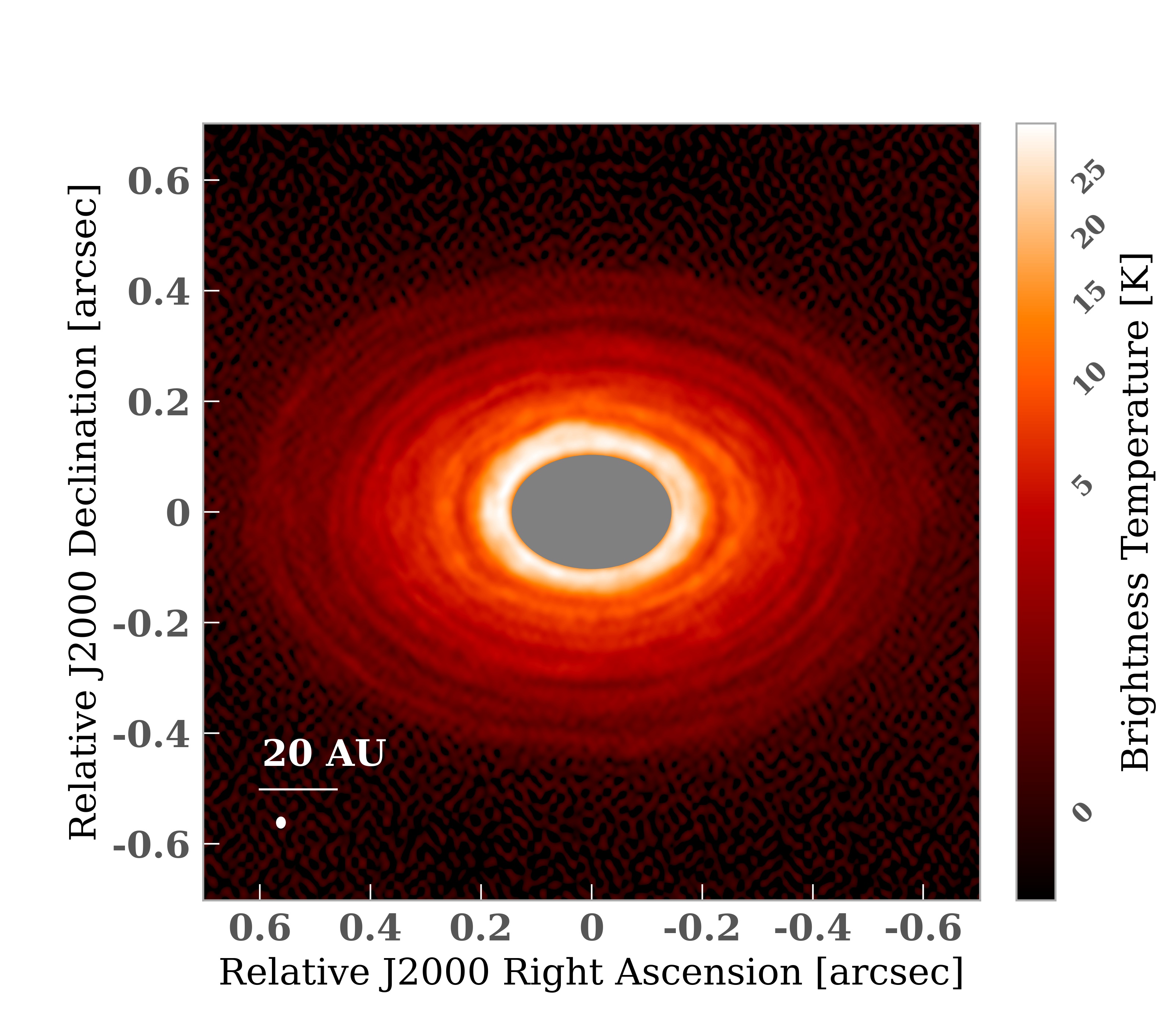}
\caption{Simulated 40-hour long ALMA observations for the dust emission at 1.3 mm for our disk model with an inclination of 45$^{\circ}$. The resultant RMS noise and the synthesized beam, which is shown in the lower left corner, correspond to about 1.3 $\mu$Jy/beam and 18 $\times$ 15 mas, respectively.} 
\label{fig:rms_1pt3microJy}
\end{figure}

\begin{figure}[t!]
\centering
\includegraphics[width=\columnwidth]{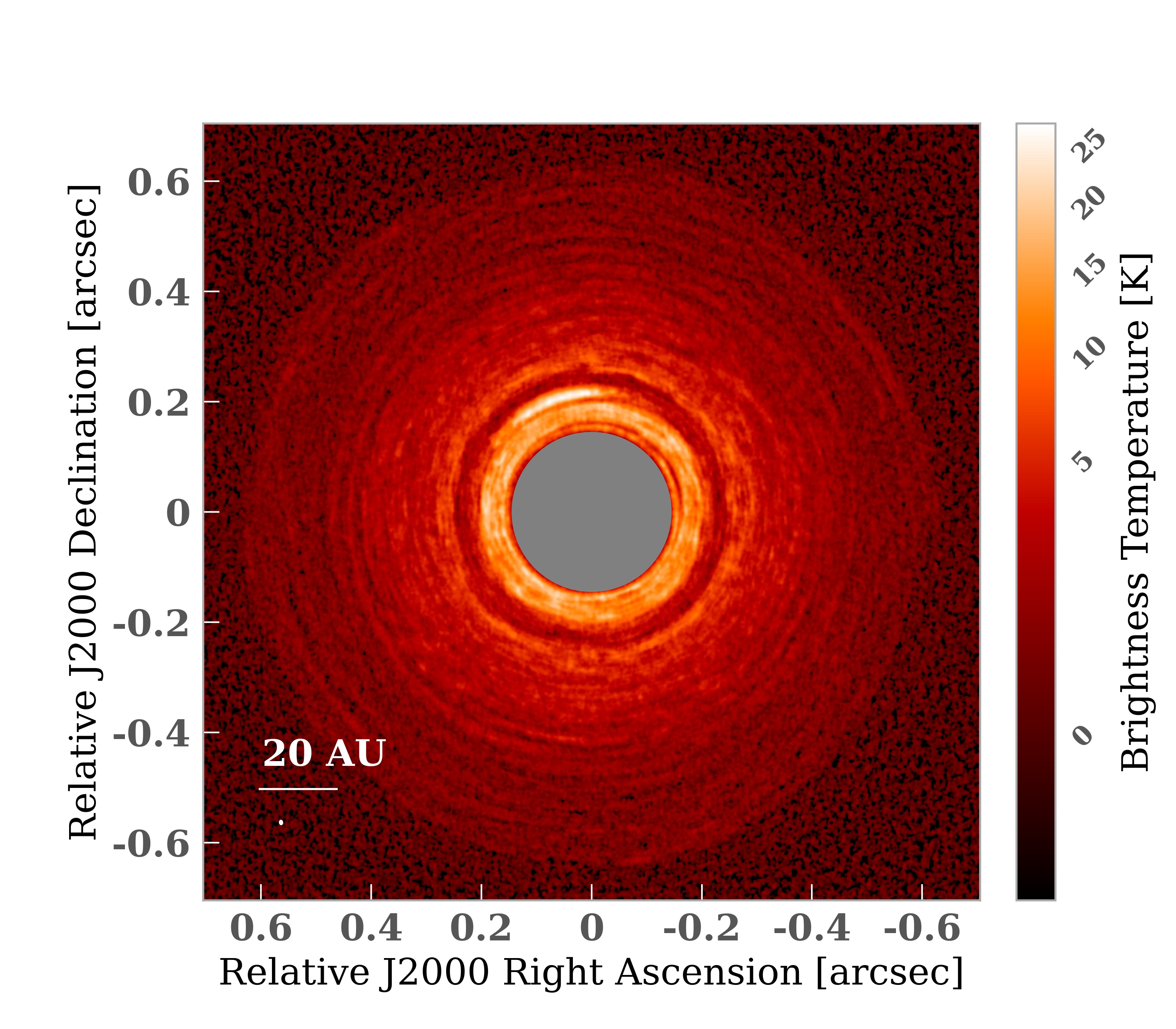}
\caption{Simulated 80-hour long ngVLA observations for the dust emission at 3 mm for our face-on disk model. The resultant RMS noise and the synthesized beam, which is shown in the lower left corner, correspond to about 0.1 $\mu$Jy/beam and 8 $\times$ 6 mas, respectively.}
\label{fig:rms_pt1microJy}
\end{figure} 

As suggested by the results presented in the previous section, the small percentage of disks showing azimuthally asymmetric features in the DSHARP sample may be in account of the observational limitations rather than of the absence of those substructures.
In this section, we investigate the potential to detect and resolve all the structures predicted from our VSI model through high-resolution observations with integration times longer than in the simulations presented above.

Figure~\ref{fig:rms_1pt3microJy} displays the result of $\sim$ 40 hour ALMA observations of our 45$^{\circ}$ inclination disk model at 1.3 mm with the ALMA longest baselines. The asymmetric bump still appears attached to the ring at the inner edge of the gap, but the much lower noise on the map allows for the detection of several arcs in the disk outer regions. Note that some of these detected features are the combination of narrower arcs blended together by the ALMA resolution beam (see top panels in Fig.~\ref{fig:all_45_deg}).
This demonstrate that ALMA observations with longer integration times than those conducted so far for nearby protoplanetary disks have the potential to unveil some of these small-scale structures predicted by the VSI model.

In Figure~\ref{fig:rms_pt1microJy} we display the image obtained from 80-hour long observations at 3 mm with the ngVLA for our face-on disk model. 
This 1 au-resolution image with rms noise of $0.1~\micron/$beam at 3 mm would detect and separate several narrow arcs in both the outer disk regions and in the inner ring, and would also resolve with several resolution beams both the gap structure at 34 au as well as the bump on the inner edge of the gap. 

\subsection{On the observability of these structures in disks in nearby star forming regions}

The total flux densities of the disk model presented in this work (Table~\ref{table:obs}) are  similar to the observed fluxes of the disks targeted by the DSHARP program. Among the 20 targets observed at 1.25 mm from that survey, 9 are brighter than our model, 8 of which have an estimated stellar mass lower than $1~M_{\odot}$. Since these disks are in the bright tail end of the observed disk luminosity in nearby star forming regions, one may wonder how many disks which have similar or higher fluxes than our disk model are present in these regions. For these systems, observations with characteristics similar to those presented in this work have the potential to detect the same substructures with similar, or higher, signal-to-noise ratios than obtained here.

We note here that the comparison between disk fluxes in this section is done assuming a face-on inclination for our disk model, whereas 
models with higher inclinations have lower fluxes. At the same time, since our models do not include dust at stellocentric radii lower than 20 au, an extension to regions closer to the star would increase the predicted fluxes; for example, under the assumptions of optically thick emission and power-law radial profile for the dust temperature $T(r) \propto r^{-0.5}$ at $r < 20$ au, the total flux of the disk would increase by $\approx 40\%$.


Using the Submillimeter Array (SMA) at a wavelength of 1.3 mm \citet{andrews13} performed a survey of 210 disks in Class II YSOs in the Taurus star forming region. Out of this sample, 9 sources have a flux density higher 150 mJy, the flux density of our model at 1.3 mm while 22 of them have a flux density higher than 75 mJy at the same wavelength. 
A search in the data archive for the Combined Array for Research in Millimeter-wave Astronomy (CARMA) array find 14 out of 149 disks in Taurus with a flux density higher than 16 mJy at 3 mm, which corresponds to the flux density of our model at that wavelength, and 35 disks with a flux density higher than 8 mJy.

Another star forming region at a similar distance of $\sim$ 137 pc is Ophiuchus \citep{ortiz17}, which contains more than 300 disks. As part of the \textit{Ophiuchus DIsc Survey Employing ALMA} (ODISEA),  \citet{cieza19} studied a sample of 147 young stars in the Ophiuchus molecular cloud at 1.3 mm. Out of this sample, 7 disks are found with a flux density higher than 150 mJy, and 14 with a flux density higher than 75 mJy.

\subsection{Measuring the proper motion of the dust bump with the ngVLA}

We explore here the potential of the ngVLA to measure the orbital proper motion of the azimuthal bump in the dust emission. Detecting the orbital motion of this structure, along with any other compact source of emission in the disk, would be important to distinguish it from other possible background or foreground sources which could be detected at the high sensitivity of the ngVLA. 

Given the angular resolution of the ngVLA at 3 mm and the signal-to-noise ratio of the bump, the ngVLA would be able to detect at $3\sigma$ the proper motion due to Keplerian rotation with observations separated by only $3-4$ weeks.
Multi-epoch observations separated by a slightly longer time period would enable to rule out the hypothesis that this structure, or other similar compact ones, are from sources not belonging to the disk, or produced by and corotating with a planet lying closer to or further from the star~\citep[see][]{Ren:2020}.




\section{Conclusion}
\label{sec:conclusion}

In this work we presented the observational predictions for the dust thermal continuum emission at sub-mm to cm wavelengths from high-resolution 3D radiative hydrodynamical models which trace the evolution of gas and dust grains of different sizes in young circumstellar disks \citep{Flock:2020}. Our model accounts for hydrodynamical processes in the gas component, and specifically predicts the development of VSI in the disk regions at stellocentric radii larger than $20-25$ au. Our model indicates that this hydrodynamic instability can strongly affect the dynamics of grains with different sizes. 

In this study we characterized the substructures expected in the dust emission at sub-mm to cm wavelengths, and investigated the capabilities of current (ALMA) and future (ngVLA) interferometers to detect and resolve these substructures in nearby disks, and test the predictions of disk models undergoing VSI. 

Our main findings are summarized as follows: 
\begin{itemize}
    \item The most prominent dust substructure predicted by our VSI model is an azimuthally symmetric gap at about 34 au from the central star, in correspondence to the stellocentric radius at which VSI starts to become effective. The local radial pressure maxima at the inner and outer edges of the gap produce bright rings that are visible at all the wavelengths investigated here, i.e. between 0.87 mm and 1 cm. We show that the gap structure would be marginally resolved by ALMA with the observational characteristics of the DSHARP program, and is consistent with the ``plateau''-like features observed in some of the disks in that survey. Possible future observations with the ngVLA have the potential to resolve these structures with several resolution beams across, especially at 3 mm. 
    \item Towards the inner edge of the gap at 30 au from the star, a significant azimuthally asymmetric bump-like substructure is seen in the model synthetic images at all wavelengths, with contrast from the neighboring regions increasing at longer wavelengths. This  pattern is due to the dust size-dependent mechanism of dust trapping in a gaseous vortex formed at the inner edge of the gap.
    We show that ALMA DSHARP-like observations at 1.3 mm cannot detect this azimuthal structure. This suggests that the low fraction of detected azimuthally asymmetric structures from current ALMA observations may be the result of observational bias due to limitations in angular resolution. High angular resolution observations, preferably at longer wavelengths, e.g. with a future ngVLA, are required to detect and characterize these substructures.
    \item The VSI model predicts the presence of more subtle radially narrow arcs in the fainter outer regions of the disk, as well as in the dust ring at the inner edge of the gap. We estimate that ALMA observations of young disks with longer integration times that those performed so far have the potential to detect these substructures in the disk outskirts. Their detection would provide evidence for VSI acting in real systems. Similarly long integration observations with the ngVLA can resolve also the radially narrower arcs in the inner ring.
    \item Observations of edge-on disks with VSI can resolve the relatively large vertical extent of the dust emission expected in regions where VSI is effective; the wavelength dependence of the vertical extent of the emission can test the size-dependent vertical stirring expected for dust grains. 
    \item Given the high astrometric precision expected for the ngVLA, multi-epoch observations separated by about a month can detect the orbital motion of azimuthal structures such as the dust bump at its location in the disk. This would enable the confirmation that this structure belongs to the disk rather than being a background or foreground source. Detecting orbital motion near the local Keplerian speed would also rule out the possibility that a feature was produced by and corotating with a planet lying closer to or further from the star.
    
\end{itemize}

This study shows that disk models with the development of the vertical shear instability can produce dust substructures which can be observed with sub-mm and radio interferometers. In fact, some of the dust substructures already observed in nearby disks with ALMA are consistent with the predictions of our VSI disk model. 

Future ALMA observations with longer integration times than those already performed have the potential to unveil radially narrow arc features that are due to dust trapping in VSI-effective regions of the disk, at least in the brightest disks in nearby star forming regions.  

A future ngVLA has the potential to detect and resolve even more predicted substructures down to $\sim 1$ au resolution, and together with ALMA can investigate the wavelength dependence of the dust emission. This would foster the observational characterization of regions of local dust accumulation in young disks, a process that is currently thought to be necessary for the formation of planetesimals in these systems.

\acknowledgements 
We thank the anonymous referee for his/her helpful comments. 
This work was
supported in part by the ngVLA Community Studies program, coordinated by the National Radio Astronomy Observatory,
which is a facility of the National Science Foundation operated
under cooperative agreement by Associated Universities, Inc.
Parts of the research were carried out at the Jet Propulsion Laboratory, California Institute of Technology, under contract 80NM0018D0004 with the National Aeronautics and Space Administration.
The contributions of DB, LR, and NT were supported in part by JPL's HBCU-MSI Program and by NASA's Exoplanets Research Program through grant 17-XRP17$\_$2-0081.  M.F. acknowledges funding from the European Research Council (ERC) under the European Union’s Horizon 2020 research and innovation program (grant agreement No. 757957).

\software{RADMC-3D \citep{Dullemond:2012} \& CASA \citep{McMullin:2007}}

\bibliographystyle{aasjournal}
\begin{singlespace}
\bibliography{vsi_bib}
\end{singlespace}

\appendix

\section{Distribution of dust in grain size bins}
\label{sec:appendix_dust_bins}

To distribute the dust mass in the three size bins we follow a similar approach as described in \citet{Ruge:2016}.
In the case of the two bins with larger dust grains, i.e. the 100 $\micron$ and 1 mm grains, each of the particles of radius $a_g$ represents a certain number of dust particles $\tilde{N}_g(a_g)$.
The initial total dust mass $M_\text{d}$ amounts to $1\%$ of the total gas mass $M_\text{gas}$. We calculate the total dust mass directly from the gas mass in our simulation domain.
The total dust mass is the sum of the mass of the small dust, $M_\text{sd}$, and of the larger dust, $M_\text{ld}$, which incorporates the 100 $\micron$ and the 1 mm grain size bins:  $M_\text{d} = M_\text{sd} + M_\text{ld}$. 

To calculate the amount of mass in the small and large dust we have to first assume a grain size distribution.
The grains are homogeneous, compact and spherical, and follow the size distribution
\begin{equation}                                                                                                                                                         
 \cfrac{\text{d}n(a)}{\text{d}a} \propto {a}^{-3.5}.\label{eq:grainsizedist}                                                                                             
\end{equation}
For the small grains we assume grains between 0.1 and 10 micron in size, and we use ten bin sizes. The number ten here is arbitrary as it is only used to determine the dust fraction in the small size bin. 
The mass $M_\text{sd}$ can be then expressed in the following way:
\begin{equation}                                                                                                                                                         
  M_\text{sd} = \cfrac{4}{3}\, \pi \, \rho_\text{md} \sum_{i=1}^{10} a_i^3 \cdot N(a_i),                                                                               
\end{equation}
where $\rho_\text{md} = 2.7\, \rm g cm^{-3}$ is the density of the dust material and $N(a_i)$ is the absolute number of particles with radius $a_i$. Following the grain\
 size distribution, $N(a_i)$ can be written as:
\begin{equation}                                                                                                                                                         
 N(a_i) = N_\text{max} \cdot \left(\cfrac{a_i}{a_\text{max}}\right)^{-2.5}\label{glg:nai},                                                                               
\end{equation}
with the total amount of particles $N_\text{max}$ with radius $a_\text{max}$.
The mass of the large dust grains can be described in the same way:
\begin{equation}                                                                                                                                                         
  M_\text{ld} = \cfrac{4}{3}\, \pi \, \rho_\text{md} \sum_{g=1}^{10} a_g^3 \cdot N(a_g).\end{equation}
Because of Eq. \ref{glg:nai}, we finally get\begin{equation}                                                                                                             
M_\text{sd} + M_\text{ld} = \cfrac{4}{3}\, \pi \, \rho_\text{md} \cfrac{N_\text{max}}{a_\text{max}} \left(\sum_{i=1}^{10} a_i^{0.5} + \sum_{g=1}^{2} a_g^{0.5}\right)\label{glg:staubmasse_flock}.                                                                                                         
\end{equation}
Through reordering of Eq. \ref{glg:staubmasse_flock}, $N_\text{max}$ -- and therefore every $N(a_i)$ and $N(a_g)$ -- can be calculated. Finally, $\tilde{N}_g$ is given by 
\begin{equation}                                                                                                                                                         
 \tilde{N}_g = \cfrac{N(a_g)}{n_{bin}},                                                                                                                                  
\end{equation}
with the $n_{bin}=500000$ being the number of particles per grain size.
To calculate the dust surface density we simply  integrated over a radial bin size $\Delta r$ and then calculated the given dust mass per unit area: \begin{equation}     
\Sigma_{\rm{dust}}= \frac{\int_{\Delta r} M_\text{d}\, dr} {\int_{\Delta r} \int_\phi r\, dr\, d\phi}.                                                                        
\label{glg:surface_density}                                                                                                                                              
\end{equation}
For the given distribution we obtain a dust-to-gas mass ratio of $M_\text{sd}/M_g=0.0033$, $M_{\rm{100 \mu m}}/M_g=0.0016$ and $M_{\rm{1 mm}}/M_g=0.0051$.

\section{Radial Profiles of the Model Surface Brightness}
\label{sec:appendix_radial_profiles}
\begin{figure*}[thb!]
\centering
\includegraphics[width=\textwidth]{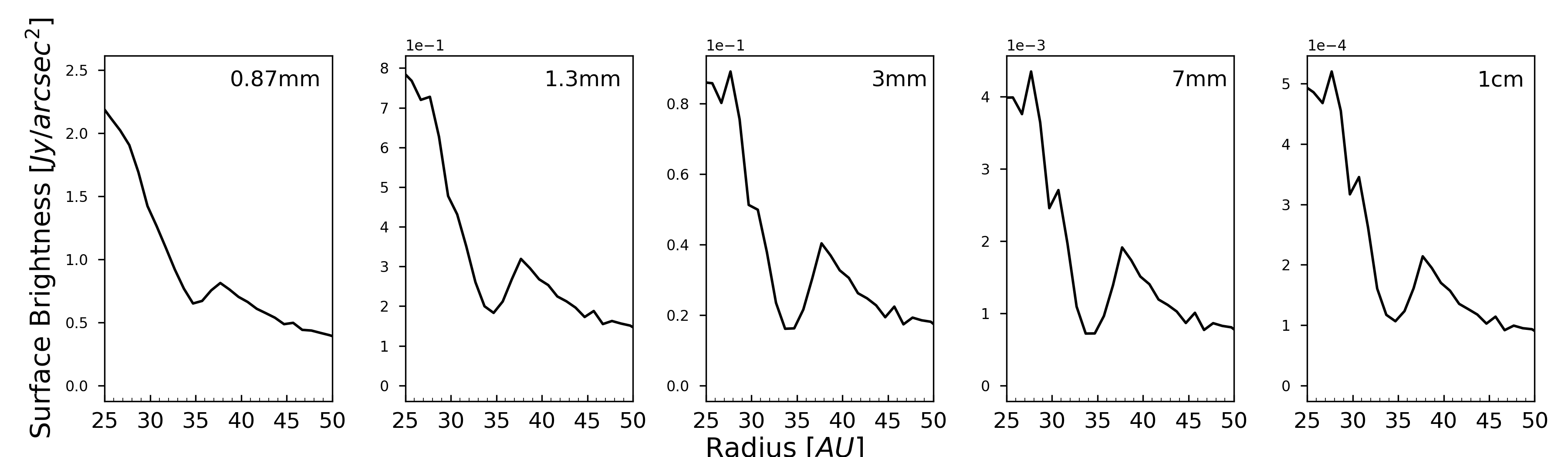}
\caption{Radial profiles of dust continuum surface brightness for the face-on model at the wavelengths specified in each panel. Each radial profile has been derived after averaging over the azimuthal coordinate.}
\label{fig:mod_rad_prof}
\end{figure*}

Figure~\ref{fig:mod_rad_prof} displays the radial profiles of the surface brightness for the dust continuum emission of our models at different wavelengths. Some of the key properties of the predicted gap, such as radial width and depth, vary with wavelength. This is a consequence of the size-dependent coupling between dust grains and gas, and the fact that the dust emission at different wavelengths is dominated by grains with different sizes. Additionally, the dependence on wavelength of the optical depth also plays a role on the emission of these structure, as regions with low optical depth are more sensitive to the dust density, and its spatial variation, than regions with higher optical depth.  
Figure~\ref{fig:mod_rad_prof} shows that the peak of the dust emission at the outer edge of the gap is higher than in the gap by factors of $\approx 1.2\times$ at 0.87 mm, $\approx 1.6\times$ at 1.3 mm, and $\approx 2\times$ at 3, 7 mm and 1 cm, respectively. 

\section{Azimuthal Profiles of the Model Surface Brightness} 
\label{sec:appendix_azimuthal_profiles}
\begin{figure*}[thb!]
\centering
\includegraphics[width=\textwidth]{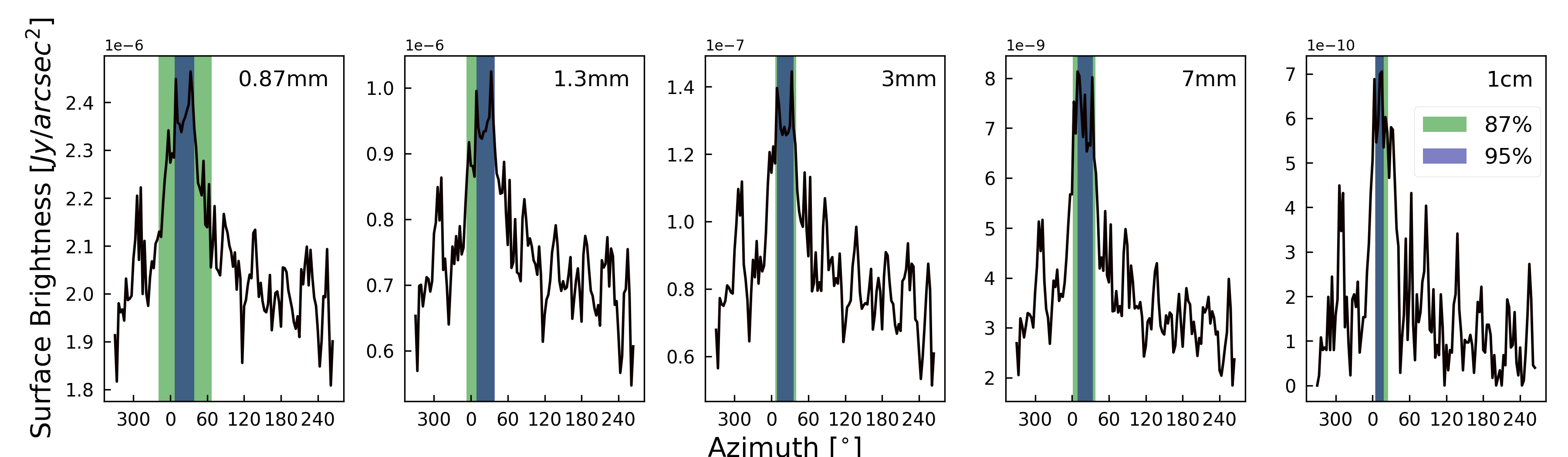}
\caption{
Azimuthal profiles at the radial location of the bump for the surface brightness of the dust continuum emission at different wavelengths as specified in each plot. The horizontal widths of the green and blue vertical bars indicate the full widths at 87$\%$ and 95$\%$ from the peak emission, respectively.}
\label{fig:az_plt}
\end{figure*} 

The wavelength-dependence of the dust emission in the region around the bump structure is highlighted in Figure~\ref{fig:az_plt}, which displays the azimuthal profiles of the dust surface brightness at the radial location of the bump. Overlaid on each graph are the full width azimuthal intervals drawn at 87$\%$ and 95$\%$ of the peak emission from the bump. For increasing wavelengths from 0.87 mm to 1 cm, the azimuthal intervals at 87$\%$ of the peak emission are 86$^{\circ}$, 45$^{\circ}$, 32$^{\circ}$, 42$^{\circ}$, and 21$^{\circ}$, respectively. The azimuthal intervals at 95$\%$ of the peak are 32$^{\circ}$, 30$^{\circ}$, 27$^{\circ}$, 25$^{\circ}$, and 13$^{\circ}$, respectively. The decrease with wavelength of the azimuthal widths at 87$\%$ and 95$\%$ can be fit with power-law functions with power-law coefficients of about $-0.4$ and $-0.3$, respectively. 

Also the morphology of the azimuthal bump between azimuthal angles of $110^{\circ}$ and $120^{\circ}$ shows some variation across the different wavelengths. 
At 0.87 mm, 1.3 mm, and 3 mm, the profiles also show a notable local minimum in emission around the central region of the dust trap, and this minimum becomes less apparent at longer wavelengths.

The relative contrast between the dust emission in the bump and that of the outer ring also increases with wavelength, similarly to the azimuthally symmetric structures discussed in the previous section. As for the discussion on the azimuthal extent of the bump at different wavelengths, this pattern is in line with the spatial segregation of grains with different sizes towards the center of the putative gaseous vortex.

\end{document}